\documentclass[11pt,draftcls,onecolumn]{IEEEtran}
\usepackage{amsfonts,bm,amsmath,comment,color,cite,amssymb,subfigure,multirow}
\usepackage[]{graphicx}
\usepackage[linesnumbered,ruled]{algorithm2e}

\def\bzero{{\mathbf{0}}}
\def\bone{{\mathbf{1}}}
\def\c1{{\textcircled{a}}}
\def\ba{{\mathbf{a}}}
\def\bb{{\mathbf{b}}}

\def\bd{{\mathbf{d}}}

\def\bh{{\mathbf{h}}}

\def\bm{{\mathbf{m}}}
\def\bn{{\mathbf{n}}}

\def\br{{\mathbf{r}}}
\def\bs{{\mathbf{s}}}
\def\bt{{\mathbf{t}}}
\def\bu{{\mathbf{u}}}

\def\bx{{\mathbf{x}}}
\def\by{{\mathbf{y}}}

\def\bA{{\mathbf{A}}}
\def\bB{{\mathbf{B}}}
\def\bC{{\mathbf{C}}}
\def\bD{{\mathbf{D}}}
\def\bE{{\mathbf{E}}}

\def\bG{{\mathbf{G}}}
\def\bH{{\mathbf{H}}}
\def\bI{{\mathbf{I}}}

\def\bK{{\mathbf{K}}}
\def\bL{{\mathbf{L}}}

\def\bM{{\mathbf{M}}}

\def\bR{{\mathbf{R}}}
\def\bS{{\mathbf{S}}}
\def\bT{{\mathbf{T}}}

\def\bX{{\mathbf{X}}}
\def\bY{{\mathbf{Y}}}

\def\bzero{{\mathbf{0}}}
\def\bone{{\mathbf{1}}}

\def\txn{{\textrm{n}}}
\def\txh{{\textrm{h}}} % short for noise
\def\tr{{\textrm{tr}}}

\def\HH{{\dagger}}
\renewcommand\Re{{\textrm{Re}}}
\renewcommand\vec{{\textrm{vec}}}

\begin{document}
\title{Constrained Radar Waveform Design for Range Profiling}
\author{Bo~Tang,~%,~\IEEEmembership{Member,~IEEE,}
        Jun~Liu,~\IEEEmembership{Senior Member,~IEEE,}
        Hai~Wang,
        and~Yihua~Hu
        %and~Jane~Doe,~\IEEEmembership{Life~Fellow,~IEEE}% <-this % stops a space
\thanks{Bo Tang, Hai~Wang, and Yihua~Hu are with the College of Electronic Engineering, National University of Defense Technology, Hefei 230037, China (email: tangbo06@gmail.com; wanghai17@nudt.edu.cn;yh\_hu@263.net).}% <-this % stops a space
\thanks{Jun Liu is with the Department of Electronic Engineering and Information Science, University of Science and Technology of China, Hefei, China (email: jun\_liu\_math@hotmail.com).}
\thanks{This work of Bo Tang was supported in part by the National Natural Science Foundation of China under Grant 61671453 and 61801500, in part by the Young Elite Scientist Sponsorship Program under Grant 17-JCJQ-QT-041, and in part by the Anhui Provincial Natural Science Foundation under Grant 1908085QF252.}
\thanks{The work of Jun Liu was supported in part by the National Natural Science Foundation of China under Grant 61871469 and in part by the Youth Innovation Promotion Association of the Chinese Academy of Sciences under Grant CX2100060053.}
\thanks{The work of Yihua Hu was supported in part by the National Natural Science Foundation of China under Grant 61871389 and in part by the Research Plan Project of the National University of Defense Technology under Grant ZK18-01-02.}
}

\markboth{Preprint}%
{Shell \MakeLowercase{\textit{et al.}}: Bare Demo of IEEEtran.cls for Journals}

\maketitle

\begin{abstract}
%\boldmath
Range profiling refers to the measurement of target response along the radar slant range. It plays an important role in automatic target recognition. In this paper, we consider the design of transmit waveform to improve the range profiling performance of radar systems. Two design metrics are adopted for the waveform optimization problem: one is to maximize the mutual information between the received signal and the target impulse response (TIR); the other is to minimize the minimum mean-square error for estimating the TIR. In addition, practical constraints on the waveforms are considered, including an energy constraint, a peak-to-average-power-ratio constraint, and a spectral constraint. Based on  minorization-maximization, we propose a unified optimization framework to tackle the constrained waveform design problem. Numerical examples show the superiority of the waveforms synthesized by the proposed algorithms.% over the widely used linear frequency-modulated  waveforms.
\end{abstract}


\section{Introduction}
Radar range profiling refers to the generation of high resolution range profiles (HRRP) by processing the target returns. The generated HRRP can be used for image formation \cite{Wehner1994highResolutionRadar} and automatic target recognition \cite{tait2005introduction} in radar systems. Conventional wideband radar systems usually transmit high-range-resolution waveforms (e.g., the linear frequency-modulated (LFM) waveforms, also called chirp waveforms) and use matched filtering (MF) to obtain range profiles. However, the optimality of transmitting such waveforms is not guaranteed. Moreover, it is well-known that the MF is optimal only in the sense of maximizing signal-to-noise-ratio (SNR) for point-like target detection in the presence of white noise.

Essentially, radar range profiling is a process of extracting information from target echoes. It can also be viewed as a parameter estimation problem, where the target impulse response (TIR) is the parameter of interest. Therefore, estimators other than the MF can be used to improve the range profiling performance. Well-known data-independent estimators include mismatched filter (MMF) and least-square (LS) estimator (see, e.g., \cite{levanon2005cross,Stoica2008IV,Song2000Estimation} and the references therein). The MMF improves the signal-to-clutter-plus-noise ratio (SCNR) but suffers some SNR loss. In \cite{blunt2006APC}, the authors pointed out that, an improper selection of the processing window for the LS estimator degraded the range profiling performance. To overcome the limitation of the LS estimator, they proposed a data adaptive approach, called adaptive pulse compression (APC). The APC algorithm achieved low estimation errors and  was capable of unmasking weak targets. In \cite{Kikuchi2017APC}, this algorithm was tested on measured data and showed the superiority over the conventional MF. Moreover, if the TIR is sparse (i.e., the number of non-zero component of the TIR is small), algorithms based on sparse reconstruction can be applied to estimate the TIR (see, e.g., \cite{yardibi2010IAA} and the references therein).

In addition to deriving estimators (in the receiver), there are also considerable interests in designing waveforms (in the transmitter) in recent years (see, e.g., \cite{He2012WaveformBook,gini2012waveform,Cui2020Waveformbook} and the references therein). In \cite{Stoica2009CAN,Song2015MM}, the authors proposed several computational approaches to minimize the integrated sidelobe level (ISL) of the transmit waveforms. Indeed, if MF is used in the receiver, waveforms with low ISL are useful to suppress clutter from the neighborhood range bins. Alternatively, if MMF is used, the detection performance of radar systems can be further enhanced by jointly designing the transmit waveform and  receive filter (see, e.g. \cite{Garren2002matchedillumination,Cheng2017PolarimetricDesign,Ciuonzo2015Intrapulse,my2016TxRx,my2020Polyphase} and the references therein). However, waveforms designed for enhancing the target detection performance might not be suitable for range profiling.

For the range profiling problem, a widely used metric for designing waveforms  is  to maximize the mutual information between the TIR and the received signal (see, e.g., \cite{Bell1993IT,Leshem2007Information,Romero2011MI,Huang2015spectrum,Bica2016MI, Palama2019matched,my2019Spectrally} and the references therein). In \cite{Bell1993IT}, the author derived the optimal ``estimation" waveform maximizing the mutual information.  Results showed that the optimal estimation waveforms admitted a water-filling solution. In \cite{Leshem2007Information}, the authors extended the work in \cite{Bell1993IT}, and proposed a waveform design algorithm for estimating multiple extended targets with a phased-array radar. In \cite{Romero2011MI}, the waveform design problem in the presence of signal-dependent interference was considered. In \cite{Huang2015spectrum,Bica2016MI, Palama2019matched}, radar waveform design for spectrum sharing with communication systems was addressed. However, the waveforms synthesized by the algorithms in these works are not constant-modulus, which makes them difficult to implement in practical radar systems.

%Surprisingly, although MMSE has been used to design waveforms for multiple-input-multiple-output (MIMO) radars (see, e.g., \cite{Yang2007MMSE,my2012MMSE,Herbert2018MMSE,Herbert2018Adaptive,my2019MMSE} and the references therein), waveform design based on minimizing the MMSE for SISO radars  has not been discussed (to the best of the authors' knowledge).

In this paper, we consider the design of practically constrained waveform for range profiling with single-input-single-output (SISO) radars (which means that we extend the point-target formulation to distributed targets). Given that range profiling is a parameter estimation problem, we first consider the widely used mutual information criterion (i.e., designing waveforms based on maximizing the mutual information between the received signal and the TIR). Different from the approaches in \cite{Bell1993IT,Leshem2007Information,Romero2011MI}, which optimized the (continuous) spectrum of the waveforms, we consider the optimization of the discrete-time waveforms. This facilitates enforcing practical constraints on the waveform.  To investigate the estimation performance of the waveform designed based on mutual information maximization, we then consider the waveform design based on minimizing MMSE, and compare their performance. We develop a unified optimization framework based on minorization-maximization (MM) to tackle the encountered (non-convex) waveform design problems. Numerical examples show that the waveforms synthesized by the proposed algorithms outperform their counterparts.

%we consider two design metrics for the waveform optimization problem: one is the widely used mutual information criterion (i.e., designing waveforms based on maximizing the mutual information between the received signal and the TIR); the other is to minimize the MMSE of TIR estimation. Different from the approaches in \cite{Bell1993IT,Leshem2007Information,Romero2011MI}, which optimized the (continuous) spectrum of the waveforms, we consider the optimization of the discrete-time waveforms. Moreover, we impose several practical constraints on the waveform, including an energy constraint, a peak-to-average-power-ratio (PAPR) constraint, and a spectral constraint. To tackle the non-convex waveform design problem, we develop a unified optimization framework based on minorization-maximization (MM). Numerical examples show that the waveforms synthesized by the proposed algorithms outperform the conventional LFM waveforms.

The rest of this paper is organized as follows. Section \ref{sec:SigModel} establishes the signal model and formulates the waveform design problem.
Section \ref{sec:MinorizerConstruct} proposes an MM-based algorithm framework to tackle the non-convex waveform design problem. Section \ref{sec:QuadProblemSolve} gives methods to efficiently tackle the quadratic programming problem encountered at each iteration of the proposed MM algorithms. Section \ref{sec:Discussion} analyzes the convergence and the computational complexity of the proposed algorithms.
Section \ref{sec:Example} provides numerical examples to demonstrate the performance of the proposed algorithms.
Finally, we conclude the paper in Section \ref{sec:Conclusion}.

\emph{Notations}:
Throughout this paper, matrices are denoted by bold uppercase letters and vectors are denoted by bold lowercase letters.
$\mathbb{C}^{m\times n}$ and $\mathbb{C}^{k}$ are the sets of ${m\times n}$ matrices and $k\times1$ vectors with complex-valued entries.
$\bI$, $\bone$, and $\bzero$  denote the identity matrix, the matrix of ones, and the matrix of zeros, with the size determined by the subscript or from the context. %$\bone_{M\times N}$ and $\bzero_{M\times N}$ denote the $M\times N$ matrices of ones and zeros.
Superscripts $(\cdot)^{T}$, $(\cdot)^*$, and $(\cdot)^{\HH}$ denote the transpose, the conjugate, and the conjugate transpose.
%$\textrm{tr}(\cdot)$ stands for the trace of a square matrix.
The symbols $\det(\cdot)$ and $\textrm{tr}(\cdot)$ indicate the determinant and the trace of a square matrix.
$\|\cdot\|_2$ and $\|\cdot\|_{\infty}$ denote the Euclidean norm and the $\ell_{\infty}$-norm of a vector argument.
$\|\cdot\|_{\textrm{F}}$ denotes the Frobenius norm of a matrix argument.
$\otimes$ and $\circledast$ represent the Kronecker product and the operator of convolution.
%$\|\bx\|_2$ denotes the Euclidian norm of the vector $\bx$.
${\vec}(\bX)$ indicates the vector obtained by column-wise stacking of the entries of $\bX$.
%$\textrm{Diag}(\bx)$ denotes the diagonal matrix with the diagonal elements formed by the vector $\bx$.
%$\textrm{diag}(\bX)$ denotes the vector formed by the diagonal elements of $\bX$.
%For the diagonal matrix $\bSig^{\downarrow}$ ($\bSig^{\uparrow}$),  the superscript $\downarrow$ ($\uparrow$) means that the diagonal elements are in decreasing (increasing) order.
%$\bA(:,k)$ denotes the $k^{th}$ column of the matrix $\bA$.
$\Re(\bX)$ denotes the real part of the matrix $\bX$ (element-wise).
$\arg(x)$ represents the argument of $x$.
$\lfloor  x \rfloor$ denotes the nearest integer less than or equal to $x$.
%$\textrm{BlkDiag}([\bA;\bB])$ denotes the block-diagonal matrix formed by the matrices $\bA$ and $\bB$.
The notation $\bA \succ \bB $ ($\bA \succeq \bB$) means that $ \bA - \bB$ is positive definite (semi-definite).
%$f(\bA)$ is called monotonically increasing with respect to (w.r.t.) $\bA$, if $\bA_1 \succeq \bA_2$ implies $f(\bA_1) \geq f(\bA_2)$.
$\bx\sim \mathcal{CN}(\bm,\bR)$ means that $\bx$ obeys a circularly symmetric complex Gaussian distribution with mean $\bm$ and covariance matrix $\bR$.
Finally, $\mathbb{E}(x)$ denotes the expectation of the random variable $x$.

\newtheorem{Def}{Definition}
\newtheorem{lemma}{Lemma}
\newtheorem{theorem}{Theorem}
\newtheorem{Prop}{Proposition}

\section{Signal Model and Problem Formulation} \label{sec:SigModel}
\subsection{Signal Model}
\begin{figure*}[!htp]
\centering
\includegraphics[width = 0.7\textwidth]{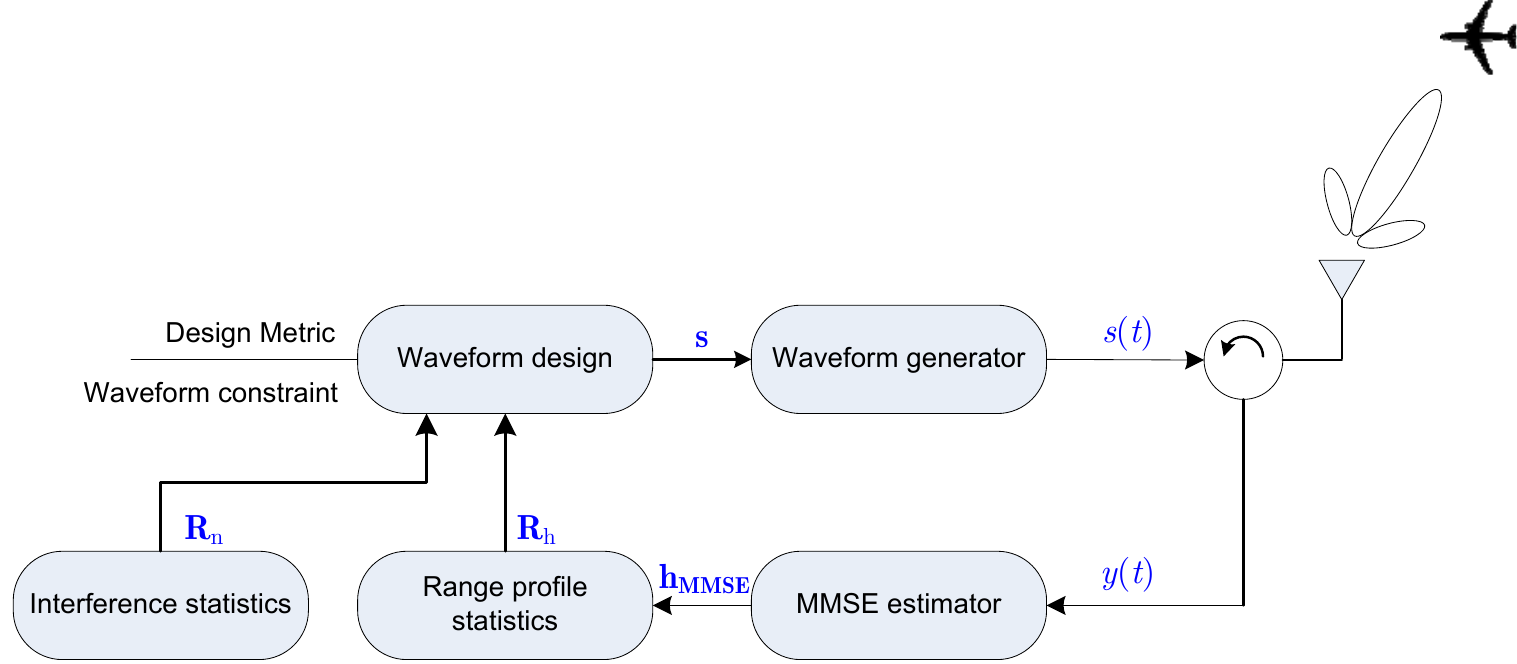}
\caption{Illustration of the radar system architecture.}
\label{Fig:System}
\end{figure*}

As illustrated in Fig. \ref{Fig:System}, we consider a wideband radar system with $s(t)$ being its baseband transmit waveform. Let $h(t)$ denote the TIR. Then the down-converted received signal can be written as
\begin{equation}%\label{}
  y(t) = \int_{\tau_1}^{\tau_2} s(\tau) h(t- \tau) \textrm{d} \tau+ n(t),
\end{equation}
where $\tau_1$ and $\tau_2$ denotes the minimum and maximum two-way propagation delays, respectively, and $n(t)$ denotes the disturbance in the receiver (accounting for possible jamming signals and receiver noise). To facilitate the following discussions, we consider the discretized signal model. Let $\bs =[s_1, s_2, \cdots, s_L]^T\in \mathbb{C}^L$ denote the (discrete-time) waveform (associated with $s(t)$), and $\bh \in \mathbb{C}^P$ denote the target response vector (associated with $h(t)$), where $L$ is the code length, and $P$ is the number of range cells with target signals \footnote{$P \approx B(\tau_2 - \tau_1)$, where $B$ is the bandwidth of the signal.}. Then the digital received signal, denoted by $\by$, can be written as
\begin{equation}\label{eq:SigModelConv}
  \by = \bh \circledast \bs + \bn,
\end{equation}
where $\bn \in \mathbb{C}^{L_0}$ denotes the samples corresponding to $n(t)$, and $L_0 = L+P-1$.

Define
\begin{align}
\bS = \left(
        \begin{array}{cccc}
         s_1    & 0      & \cdots & 0 \\
          s_2    & s_1    & \ddots & 0 \\
          \vdots & \ddots & \ddots & 0 \\
          s_L    & s_{L-1}& \cdots & 0 \\
          0      & s_L    & \ddots & 0 \\
          \vdots & \vdots & \vdots & \vdots \\
          0      & \cdots & \cdots & s_L \\
        \end{array}
      \right) \in \mathbb{C}^{L_0\times P}. \label{Eq:S}
\end{align}
Then $\by$ can be rewritten as
\begin{equation}\label{eq:SigModelMatrix}
  \by = \bS \bh + \bn.
\end{equation}

Note that similar signal models are considered in
\cite{Garren2002matchedillumination,Cheng2017PolarimetricDesign,Karbasi2015ExtendedTargets,my2016ICASSP}. In these papers, the authors assumed that the TIR is known \textit{a priori} or partially known. The assumption is justified if a template library containing TIR at all relevant target aspect angles is available, or the radar system is cognitive, i.e., the radar system can obtain the prior knowledge of the TIR from the previous estimates (see, e.g., \cite{Gurbuz2019CognitiveOverview,Palama2019matched,Horne2020AdaptiveSignalling,Smith2016cognitiveExperiments} and the references therein for a recent discussion).  Different from \cite{Garren2002matchedillumination,Cheng2017PolarimetricDesign,Karbasi2015ExtendedTargets,my2016ICASSP}, which used a deterministic TIR model, we employ a stochastic TIR model. Specifically, we assume that $\bh \sim \mathcal{CN}(\boldsymbol{\mu}_\txh, \bR_\txh)$, where $\boldsymbol{\mu}_\txh$ indicates the prior knowledge of the target response and $\bR_\txh$ accounts for the uncertainty in the prior knowledge ($\bR_\txh$ can be estimated based on multiple previous estimates, or specified by the user). In addition, we assume that the disturbance $\bn$ obeys a Gaussian distribution with zero mean and covariance matrix $ \bR_\txn$ (estimated based on target-free samples). Based on the two assumptions, we have $\by \sim \mathcal{CN}(\bS\boldsymbol{\mu}_\txh, \bR_\bs )$, where $\bR_\bs = \bS  \bR_\txh \bS^\HH + \bR_\txn$.

Next we present two metrics for designing waveforms to improve the range profiling performance.

\subsection{Waveform Design Based on Maximizing Mutual Information}
The mutual information between $\by$ and $\bh$ is given by
\begin{align}\label{eq:MI}
  {\textsf{I}}(\by;\bh)
  &= \textsf{H}(\by) - \textsf{H}(\by|\bh) \nonumber \\
  &= \log \det (\bS  \bR_\txh \bS^\HH + \bR_\txn) -  \log \det(\bR_\txn)\nonumber \\
  &= \log \det (\bR_\txn^{-1}\bS  \bR_\txh \bS^\HH + \bI),
\end{align}
where $\textsf{H}(\by)$ denotes the differential entropy of $\by$, and $\textsf{H}(\by|\bh)$ denotes the conditional differential entropy of $\by$ given $\bh$ \cite{cover2012elements}.
Thus, the waveform design problem based on maximizing mutual information can be formulated as follows:
\begin{align} \label{eq:MIproblem}
  \max_{\bs} & \ f_{{\textrm{I}}}(\bs) = \log \det (\bR_\txn^{-1}\bS  \bR_\txh \bS^\HH + \bI) \nonumber \\
  \textrm{s.t.} & \ \bs \in \mathcal{S},
\end{align}
where $\mathcal{S}$ denotes the feasibility region of $\bs$, and is determined by  the constraints on the waveforms (detailed in Subsection \ref{subsection:constraint}).
\subsection{Waveform Design Based on Minimizing MMSE}
Given the signal model in \eqref{eq:SigModelMatrix}, the MMSE estimator of $\bh$ is given by \cite{kaybook1993}
\begin{equation}\label{eq:hMMSE}
  \bh_{\textrm{MMSE}} = \boldsymbol{\mu}_h + \bR_\txh \bS^\HH (\bS \bR_\txh\bS^\HH + \bR_\txn)^{-1} (\by- \bS \boldsymbol{\mu}_h).
\end{equation}
The associated MMSE is
\begin{align}%\label{}
  \textsf{MMSE}
  &= \mathbb{E}(\|\bh - \bh_{\textrm{MMSE}}\|_2^2) \nonumber \\
  &= \tr [(\bR_\txh^{-1} +  \bS^\HH \bR_\txn ^{-1} \bS)^{-1}]  \nonumber \\
  &= \tr (\bR_\txh - \bR_\txh\bS^\HH (\bS \bR_\txh \bS^\HH + \bR_\txn)^{-1} \bS\bR_\txh),
\end{align}
where the third equality is derived using the matrix inversion lemma \cite{HornJohnson1990matrixbook}.
Thus, the waveform design problem based on minimizing MMSE is equivalent to
\begin{align}\label{eq:MMSEproblem}
  \max_{\bs} &\  f_{\textrm{E}}(\bs) = \tr (\bR_\txh\bS^\HH (\bS \bR_\txh \bS^\HH + \bR_\txn)^{-1} \bS\bR_\txh) \nonumber \\
  \textrm{s.t.} &\ \bs \in \mathcal{S}.
\end{align}

\subsection{Waveform Constraints} \label{subsection:constraint}
In this paper, we consider the following constraints:
\begin{itemize}
  \item Energy constraint. In practical radar systems, the available transmit energy for the waveform is limited. Thus, we impose the following constraint:
  \begin{equation}\label{eq:energyConstraint}
    \bs^\HH \bs \leq e_t,
  \end{equation}
  where $e_t$ is the total available transmit energy. It can be proved that the mutual information is increasing and the MMSE is decreasing with respect to (w.r.t.) the waveform energy (see Appendix \ref{Apd:A} for a proof). Thus, the optimal waveform maximizing the mutual information and the optimal waveform minimizing the MMSE must satisfy
  \begin{equation}%\label{}
    \bs^\HH \bs = e_t.
  \end{equation}
  \item PAPR constraint. To improve the efficiency of the radio frequency amplifier and avoid the nonlinear effect in the transmitter, waveforms with low PAPR is desired \cite{DeMaio2009Similarity,DeMaio2011PAPR,Stoica2012joint,Soltanalian2013crew,Wu2018MultipleConstraints,Aubry2020MultiSpectral,my2021RangeSpread}. To control the PAPR of the waveform, we consider the following constraint:
      \begin{equation}\label{eq:PAPRconstraint}
        \bs^\HH \bs = e_t, \textrm{PAPR}(\bs) \leq \rho,
      \end{equation}
      where $1\leq \rho \leq L$, and
      \begin{equation}\label{eq:DefPAPR}
        \textrm{PAPR}(\bs) = \frac{\max_{l} |s(l)|^2}{\frac{1}{L}\sum_{l=1}^{L}|s(l)|^2}.
      \end{equation}
      If $\rho = 1$, the PAPR constraint becomes the constant-modulus constraint, i.e., the constraint in \eqref{eq:PAPRconstraint} can be expressed as
      \begin{equation}\label{eq:CMC}
        |s(l)| = a_s, l=1,\cdots,L,
      \end{equation}
      where $a_s = \sqrt{e_t/L}$. If $\rho = L$, the constraint $\textrm{PAPR}(\bs) \leq \rho$ is redundant and the PAPR constraint becomes the energy constraint.
  %\item Similarity constraint. To control the shape of the ambiguity function and possess some desirable properties, we enforce the similarity constraint on the waveform:
%      \begin{equation}\label{eq:SimilarityConstraint}
%        \bs^\HH \bs = e_t, \|\bs - \bs_0\|_2^2 \leq \varepsilon,
%      \end{equation}
%      where $\bs_0$ is the reference waveform with some desirable properties, $\bs_0^\HH \bs_0 = e_t$, $0\leq \varepsilon \leq 2e_t$ is a user-specified parameter ruling the similarity region (we refer to it as the similarity parameter).
%  \item Constant-modulus and similarity constraints. The constant-modulus and similarity constraints on the waveform are as follows:
%  \begin{equation}\label{eq:CMSIM}
%    |s(l)| = a_s, l=1,\cdots,L, \|\bs - \bs_0\|_\infty \leq \varepsilon_\infty,
%  \end{equation}
%  where $\bs_0$ is the reference waveform with energy $e_t$ and modulus equal to $a_s$, and $\varepsilon_\infty$ is the similarity parameter.
  \item Spectral constraint. The increasing demand of a larger bandwidth for radar and communication systems results in a crowded spectrum and mutual interference \cite{Griffiths2015spectrum,Bockmair2019CognitivePrinciples,Carotenuto2020SpectrumManagement,yang2020MultiSpectral,Aubry2020MultiSpectral,Horne2020AdaptiveSignalling}. To improve the performance and reduce the interference for a radar operating in spectrally crowded environments, we enforce a spectral constraint on the waveform (see a similar constraint in \cite{Carotenuto2020SpectrumManagement,Aubry2014spectrally,Aubry2016spectrally,Aubry2020MultiSpectral}):
      \begin{equation}\label{eq:SpectralConstraint}
        \bs^\HH \bs = e_t, \bs^\HH \bR_\textrm{I} \bs \leq E_\textrm{I},
      \end{equation}
      where $E_\textrm{I}$ is the maximum allowed interference that can be tolerated by the communication systems, $\bR_\textrm{I} = \sum_{k=1}^{K_\textrm{rad}} w_k \bR_{\textrm{I},k}$, ${K_\textrm{rad}}$ is the number of nearby communication systems, $w_k$ is the weight corresponding to the $k$th communication system (which is a user-defined parameter and represents different emphasis on various communication systems),
      \begin{equation*}%\label{eq:RI}
        \bR_{\textrm{I},k}(m,n) = \begin{cases}
          f_2^k - f_1^k,&m=n\\
          \frac{e^{j2\pi f_2^k(m-n)} - e^{j2\pi f_1^k(m-n)}}{j2\pi(m-n)}&m\neq n,
        \end{cases}
      \end{equation*}
      $f_2^k$ and $f_1^k$ are the upper and the lower normalized frequencies of the $k$th communication system, respectively. Note that $f_2^k$ and $f_1^k$ can be obtained from spectral regulations, or by spectrum sensing (e.g., by a cognitive radar \cite{Carotenuto2020SpectrumManagement,Horne2020AdaptiveSignalling}).
\end{itemize}

\section{Minorizer Construction} \label{sec:MinorizerConstruct}
Note that the optimization problems in \eqref{eq:MIproblem} and \eqref{eq:MMSEproblem} are in general non-convex.  To tackle the non-convex optimization problems, we develop a unified optimization framework in this section. Specifically, the framework is based on MM (see, e.g., \cite{Hunter2004MM,Sun2017MM} for a tutorial introduction to the MM algorithms). The key idea of the proposed framework is to derive minorizers for  $f_{\textrm{I}}(\bs)$ and $f_{\textrm{E}}(\bs)$. In precise, the derived minorizers (denoted by $g_\textrm{O}(\bs;\bs_k)$) should satisfy that
\begin{subequations}
  \begin{align}\label{eq:Minorizer}
   g_\textrm{O}(\bs;\bs_k) &\leq f_\textrm{O}(\bs),\\
   g_\textrm{O}(\bs_k;\bs_k) &= f_\textrm{O}(\bs_k),
\end{align}
\end{subequations}
where $\textrm{O} = \textrm{I}$ or $\textrm{E}$.
\subsection{Minorizer for $f_{\textrm{I}}(\bs)$}
Note that the objective function in \eqref{eq:MIproblem} can be rewritten as
\begin{align}%\label{}
  f_{{\textrm{I}}}(\bs) = \log \det (\bR_\txh^{\frac{1}{2}} \bS^\HH \bR_\txn^{-1}\bS\bR_\txh^{\frac{1}{2}} + \bI),
\end{align}
where we have used the fact that $\det(\bI + \bA\bB) = \det(\bI + \bB\bA) $ \cite{HornJohnson1990matrixbook}.
Using the matrix inversion lemma yields \cite{HornJohnson1990matrixbook}
\begin{align}%\label{}
  (\bR_\txh^{\frac{1}{2}} \bS^\HH \bR_\txn^{-1}\bS\bR_\txh^{\frac{1}{2}} + \bI)^{-1}
  &= \bI - \bR_\txh^{\frac{1}{2}} \bS^\HH  \bR_{\bs}^{-1}  \bS \bR_\txh^{\frac{1}{2}} \nonumber \\
  &= (\bK \bM^{-1} \bK^\HH)^{-1},
\end{align}
where $\bK = [\bI_P ,\bzero_{P \times L_0}]$, and
\begin{equation}%\label{}
  \bM =
  \begin{bmatrix}
   \bI & \bR_\txh^{\frac{1}{2}} \bS^\HH \\
   \bS \bR_\txh^{\frac{1}{2}} &  \bR_{\bs}
  \end{bmatrix}.
\end{equation}
Thus, we have
\begin{equation}%\label{}
  f_{\textrm{I}}(\bs) = \log \det (\bK \bM^{-1} \bK^\HH).
\end{equation}

It follows from Lemma 1 of \cite{my2018Efficient} that $\log \det (\bK \bM^{-1} \bK^\HH)$ is a convex function of $\bM$. Since convex functions are minorized by their supporting hyperplanes \cite{Boyd2004convexBook},  we have
\begin{equation*}%\label{}
  \log \det (\bK \bM^{-1} \bK^\HH) \geq \log \det (\bK \bM_k^{-1} \bK^\HH) + \tr(\bG_k (\bM - \bM_k)),
\end{equation*}
where
\begin{equation}%\label{}
  \bG_k = -\bM_k^{-1}\bK^\HH (\bK \bM_k^{-1} \bK^\HH)^{-1} \bK \bM_k^{-1}
\end{equation}
is the gradient of $\log \det (\bK \bM^{-1} \bK^\HH)$ at $\bM_k$ (which can be verified by using the results from \cite{hjorungnes2007complex}),
\begin{equation}%\label{}
  \bM_k =
  \begin{bmatrix}
   \bI & \bR_\txh^{\frac{1}{2}} \bS_k^\HH \\
   \bS_k \bR_\txh^{\frac{1}{2}} &  \bR_{\bs,k}
  \end{bmatrix},
\end{equation}
$\bR_{\bs,k} = \bS_k  \bR_\txh \bS_k^\HH + \bR_\txn$, and $\bS_k$ is formed by $\bs_k \in \mathcal{S}$ (which is the waveform at the $k$th iteration).

Let us partition $\bG_k$ as
\begin{equation}\label{eq:PartitionG}
  \bG_k =
  \begin{bmatrix}
    \bG_k^{11} & \bG_k^{12} \\
    \bG_k^{21} & \bG_k^{22}
  \end{bmatrix},
\end{equation}
where $\bG_k^{11} \in \mathbb{C}^{P \times P}$,  $\bG_k^{12} = (\bG_k^{21})^\HH \in \mathbb{C}^{P \times L_0}$, and $\bG_k^{22} \in \mathbb{C}^{L_0 \times L_0}$. Define $\bL = [\bzero_{L_0 \times P}, \bI_{L_0}]$. Then we have
\begin{align}%\label{}
  \bG_k^{11} &= \bK \bG_k \bK^\HH = -(\bR_\txh^{\frac{1}{2}} \bS_k^\HH \bR_\txn^{-1}\bS_k\bR_\txh^{\frac{1}{2}} + \bI),\\
  \bG_k^{21} &= \bL \bG_k \bK^\HH = -\bR_\txn^{-1} \bS_k \bR_\txh^{\frac{1}{2}}, \label{Gk21}\\
  \bG_k^{22} &= \bL \bG_k \bL^\HH = - \bG_k^{21} (\bK \bM_k^{-1} \bK^\HH)^{-1} \bG_k^{12}.
\end{align}
Using the matrix inversion lemma and after some algebraic manipulations, $\bG_k^{22}$ can be simplified into
\begin{equation}\label{Gk22}
  \bG_k^{22} = \bR_{\bs,k}^{-1} - \bR_\txn^{-1}.
\end{equation}
Using the partition of $\bG_k$ in \eqref{eq:PartitionG}, we have
\begin{align}%\label{}
  \tr(\bG_k \bM) = &\tr(\bG_k^{11}) + \tr(\bG_k^{22} \bR_\txn) + 2\Re(\tr(\bS^\HH \bG_k^{21} \bR_\txh^{\frac{1}{2}}  ))\nonumber \\
  & + \tr( \bG_k^{22} \bS \bR_\txh\bS^\HH).
\end{align}
Thus, $f_{\textrm{I}}(\bs)$ is minorized by
\begin{equation}\label{eq:mutualInformationMinorizer}
  g_{\textrm{I}}(\bs;\bs_k) = c_{\textrm{I}} + 2\Re(\tr(\bS \bR_\txh^{\frac{1}{2}} \bG_k^{12} )) + \tr( \bG_k^{22} \bS \bR_\txh\bS^\HH),
\end{equation}
where $c_{\textrm{I}}= \tr(\bG_k^{11}) + \tr(\bG_k^{22} \bR_\txn) +  \log \det (\bK \bM_k^{-1} \bK^\HH) - \tr(\bG_k \bM_k)$.
\begin{Prop} \label{Prop:1}
Define $\bE = [\bE_1^T, \bE_2^T, \cdots, \bE_P^T]^T$, $\bE_p  = [\bzero_{L \times (p-1)}, \bI_L,  \bzero_{L \times (P-p)}]^T\in \mathbb{C}^{L_0 \times L}$, $p=1,\cdots,P$, then we have
\begin{equation}
  g_{\textrm{I}}(\bs;\bs_k) = c_{\textrm{I}} + 2\Re(\bs^\HH \ba_{\textrm{I},k}) + \bs^\HH \bA_{\textrm{I},k} \bs,
\end{equation}
where $\ba_{\textrm{I},k} = \bE^\HH \vec(\bG_k^{21} \bR_\txh^{\frac{1}{2}} )$,
$\bA_{\textrm{I},k} = \bE^\HH (\bR_\txh^* \otimes \bG_k^{22} ) \bE $.
\end{Prop}
\begin{IEEEproof}
  See Appendix \ref{Apd:Prop1}.
\end{IEEEproof}

According to Proposition \ref{Prop:1}, the mutual information maximization problem of an MM algorithm based on \eqref{eq:mutualInformationMinorizer} (at the $(k+1)$th iteration) can be reformulated as
  \begin{align} \label{eq:MIproblemMinorized}
  \max_{\bs} & \ g_{\textrm{I}}(\bs;\bs_k) = c_{\textrm{I}} + 2\Re(\bs^\HH \ba_{\textrm{I},k}) + \bs^\HH \bA_{\textrm{I},k} \bs \nonumber \\
  \textrm{s.t.} & \ \bs \in \mathcal{S}.
\end{align}

\subsection{Minorizer for $f_{\textrm{E}}(\bs)$}
\begin{lemma} \label{Lemma:2}
  Assume that $\bB \succ \bzero$. Then $\psi(\bA,\bB) = \tr(\bA^\HH \bB^{-1} \bA)$ is jointly convex w.r.t. $\bA$ and  $\bB$, and is minorized by
  \begin{align}\label{eq:jointlyConvex}
    \tr(\bA^\HH \bB^{-1} \bA) \geq 2 \Re(\tr(\bA_k^\HH \bB_k^{-1}\bA)) - \tr(\bB_k^{-1} \bA_k \bA_k^\HH \bB_k^{-1} \bB).
  \end{align}
\end{lemma}
\begin{IEEEproof}
See Appendix \ref{Apd:Lemma1}.
\end{IEEEproof}

Note that $\bS \bR_\txh \bS^\HH + \bR_\txn \succ \bzero$. Using Lemma \ref{Lemma:2} (by substituting $\bA$ with $\bS\bR_\txh$ and $\bB$ with $\bS \bR_\txh \bS^\HH + \bR_\txn$), we obtain
\begin{align*}%\label{}
  \tr (\bR_\txh\bS^\HH (\bS \bR_\txh \bS^\HH + \bR_\txn)^{-1} \bS\bR_\txh)
\geq &2 \Re(\tr(\bH_k^{\HH}\bS) %\nonumber\\&
- \tr(\bT_k (\bS \bR_\txh \bS^\HH + \bR_\txn) ),
\end{align*}
where $\bH_k =  \bR_{\bs,k} ^{-1} \bS_k \bR_\txh^2$, and $\bT_k = \bR_{\bs,k}^{-1} \bS_k \bR_\txh^2 \bS_k^\HH \bR_{\bs,k}^{-1} $.
%\section{SPICE}
Therefore, $f_{\textrm{E}}(\bs)$ is minorized by
\begin{equation}%\label{}
  g_{\textrm{E}}(\bs;\bs_k) = c_{\textrm{E}} + 2 \Re(\tr(\bH_k\bS)) - \tr(\bT_k \bS \bR_\txh \bS^\HH),
\end{equation}
where $c_{\textrm{E}}= -\tr(\bT_k \bR_\txn)$.
Similar to that in Proposition \ref{Prop:1}, $g_{\textrm{E}}(\bs;\bs_k)$ can be rewritten as
   \begin{equation}\label{eq:MMSEminorizer}
      g_{\textrm{E}}(\bs;\bs_k) = c_{\textrm{E}} + 2\Re(\bs^\HH \ba_{\textrm{E},k}) + \bs^\HH \bA_{\textrm{E},k} \bs,
   \end{equation}
where $\ba_{\textrm{E},k} = \bE^\HH\vec(\bH_k)$, and $\bA_{\textrm{E},k} = -\bE^\HH (\bR_\txh^* \otimes \bT_k ) \bE $.

Therefore, the MMSE minimization problem of an MM algorithm based on \eqref{eq:MMSEminorizer} (at the $(k+1)$th iteration) is equivalent to
  \begin{align} \label{eq:MMSEproblemMinorized}
  \max_{\bs} & \ g_{\textrm{E}}(\bs;\bs_k) = c_{\textrm{E}} + 2\Re(\bs^\HH \ba_{\textrm{E},k}) + \bs^\HH \bA_{\textrm{E},k} \bs \nonumber \\
  \textrm{s.t.} & \ \bs \in \mathcal{S}.
\end{align}

\section{Solving the Quadratic Programming Problem} \label{sec:QuadProblemSolve}
In this section, we propose algorithms to solve the following minorized problem (which is a constrained quadratic programming problem) at the $(k+1)$th iteration:
  \begin{align} \label{eq:ConstrainedQuadProblem}
  \max_{\bs} & \ g_{\textrm{O}}(\bs;\bs_k) = c_\textrm{O}  + 2\Re(\bs^\HH \ba_{\textrm{O},k}) + \bs^\HH \bA_{\textrm{O},k} \bs \nonumber \\
  \textrm{s.t.} & \ \bs \in \mathcal{S},
\end{align}
where $\textrm{O} = \textrm{I}$ or $\textrm{O} = \textrm{E}$.
\subsection{Energy Constraint}
The minorized problem under the energy constraint can be formulated as follows:
  \begin{align} \label{eq:QuadEnergy}
  \max_{\bs} & \ 2\Re(\bs^\HH \ba_{\textrm{O},k}) + \bs^\HH \bA_{\textrm{O},k} \bs \nonumber \\
  \textrm{s.t.} & \ \bs^\dagger \bs \leq e_t.
\end{align}

Note that $\bA_{\textrm{O},k} \preceq \bzero$ (We refer to Appendix \ref{Apd:B} for the proof). Thus, the optimization problem in \eqref{eq:QuadEnergy} is convex, meaning that its globally optimal solution can be found with polynomial time (e.g., via interior point method). In addition, we can derive a semi-closed-form expression of the optimal solution by using the method of Lagrange multipliers. To this end, let the Lagrangian associated with \eqref{eq:QuadEnergy} be
\begin{equation}%\label{}
  F(\bs,\mu) = -\bs^\HH \bA_{\textrm{O},k} \bs - 2\Re(\bs^\HH \ba_{\textrm{O},k}) + \mu (\bs^\dagger \bs - e_t),
\end{equation}
where $\mu \geq 0$ is the Lagrange multiplier associated with the energy constraint.
According to the Karush-Kuhn-Tucker (KKT) conditions \cite{Boyd2004convexBook}, the optimal solution of \eqref{eq:QuadEnergy} satisfies
\begin{align}\label{eq:KKT}
  (\mu_k \bI -  \bA_{\textrm{O},k}) \bs &=  \ba_{\textrm{O},k},\nonumber  \\
   \mu_k (\bs^\dagger \bs - e_t) &=0.
\end{align}
%Setting the derivative of the Lagrangian w.r.t. $\bs$ to zero, we can obtain that the optimal solution is given by
if $\mu_k \neq 0$, then
\begin{equation}\label{eq:solutionEnergy}
  \bs = (\mu_k \bI -  \bA_{\textrm{O},k})^{-1} \ba_{\textrm{O},k},
\end{equation}
where $\mu_k$ can be obtained by solving $\ba_{\textrm{O},k}^\dagger(\mu_k \bI -  \bA_{\textrm{O},k})^{-2} \ba_{\textrm{O},k}=e_t$. Otherwise, the optimal solution is given by $\bs = -  \bA_{\textrm{O},k}^{-1} \ba_{\textrm{O},k}$.

\subsection{PAPR Constraint} \label{Subsection:PAPR}
The minorized problem under the PAPR constraint is equivalent to
  \begin{align} \label{eq:QuadPAPR}
  \max_{\bs} & \ 2\Re(\bs^\HH \ba_{\textrm{O},k}) + \bs^\HH \bA_{\textrm{O},k} \bs \nonumber \\
  \textrm{s.t.} & \ \bs^\HH \bs = e_t, \textrm{PAPR}(\bs) \leq \rho.
\end{align}

We can also tackle the problem in \eqref{eq:QuadPAPR} by MM. To this end, we note that
\begin{equation}%\label{}
  \bs^\dagger \bA_{\textrm{O},k}\bs \geq 2\Re(\bs^\dagger\bA_{\textrm{pos},k} \bs_j) + \textrm{const}_j,
\end{equation}
where $\bA_{\textrm{pos},k} = \bA_{\textrm{O},k} - \lambda_{\min}( \bA_{\textrm{O},k})\bI$, $\lambda_{\min}( \bA_{\textrm{O},k})$ is the smallest eigenvalue of $\bA_{\textrm{O},k}$, $\textrm{const}_j = 2 \lambda_{\min}( \bA_{\textrm{O},k}) e_t - \bs_j^\dagger \bA_{\textrm{O},k} \bs_j$, and $\bs_j$ is the solution at the  $j$th (inner) iteration. As a result, the objective function in \eqref{eq:QuadPAPR} is minorized by
\begin{equation}\label{eq:PAPRminorizer}
  2\Re(\bs^\dagger \bt_{k,j}) + \textrm{const}_j,
\end{equation}
where $\bt_{k,j} = \ba_{\textrm{O},k} + \bA_{\textrm{pos},k} \bs_j$.
Thus, the maximization problem of an MM algorithm based on \eqref{eq:PAPRminorizer} is
  \begin{align} \label{eq:QuadPAPRMinorized}
  \max_{\bs} & \ 2\Re(\bs^\dagger \bt_{k,j}) \nonumber \\
  \textrm{s.t.} & \ \bs^\HH \bs = e_t, \textrm{PAPR}(\bs) \leq \rho.
\end{align}
The above optimization problem can be solved by Algorithm 2 in \cite{Tropp2005AP}. In particular, if $\rho = 1$, a closed-form expression for the optimal solution can be derived as
\begin{equation}\label{eq:QuadPAPRsolution}
  s_l = a_s \exp(j \phi_{t,l}),
\end{equation}
where $\phi_{t,l} = \arg (t_l)$, and $t_l$ is the $l$th element of $\bt_{k,j}$.

Algorithm \ref{Alg:PAPR} summarizes the procedure to tackle the PAPR-constrained optimization problem in \eqref{eq:QuadPAPR}.
\begin{algorithm}[!htp]
  \caption{ \small  Optimization algorithm for \eqref{eq:QuadPAPR}.}\label{Alg:PAPR}
  \KwIn{$\bA_{\textrm{O},k},\ba_{\textrm{O},k}, \lambda_{\min}( \bA_{\textrm{O},k}),\rho$.}
  \KwOut{$\bs_{k+1}$.}
  \textbf{Initialize:} $j= 0$, $\bs_{k,j} = \bs_{k}$, $\bA_{\textrm{pos},k} = \bA_{\textrm{O},k} - \lambda_{\min}( \bA_{\textrm{O},k})\bI$.\\
    \Repeat{convergence}{
    $\bt_{k,j} = \ba_{\textrm{O},k} + \bA_{\textrm{pos},k} \bs_j$ \\
    Update  $\bs_{k,j+1}$  by solving the optimization problem in \eqref{eq:QuadPAPRMinorized} \\
    $j=j+1$
    }
    $\bs_{k+1}$ = $\bs_{k,j}$
\end{algorithm}

\subsection{Spectral Constraint} \label{Subsection:Spectral}
The minorized problem under the spectral constraint is given by
\begin{align} \label{eq:QuadSpectral}
  \max_{\bs} & \ 2\Re(\bs^\HH \ba_{\textrm{O},k}) + \bs^\HH \bA_{\textrm{O},k} \bs \nonumber \\
  \textrm{s.t.} & \ \bs^\HH \bs = e_t, \bs^\HH \bR_\textrm{I} \bs \leq E_\textrm{I}.
\end{align}

The optimization problem in \eqref{eq:QuadSpectral} is hidden-convex. Thus, it can be solved by semi-definite relaxation followed by a rank-one decomposition \cite{Ai2011rankone}. However, the computational complexity of solving a semi-definite programming problem is high ($O (L^{4.5})$, given that the primal-dual path following
method is used). We propose using the alternating direction method of multipliers (ADMM) \cite{Boyd2011ADMM} to reduce the computational complexity (As shown in Table \ref{Table:1}, the proposed ADMM has a complexity of $O(L^3)$). To apply the ADMM algorithm to this problem, we use the variable splitting trick and introduce an auxiliary variable $\bu$ as follows:
\begin{align} \label{eq:QuadSpectralADMM}
  \min_{\bs,\bu} & \ 2\Re(\bs^\HH \bar{\ba}_{\textrm{O},k}) + \bs^\HH \bar{\bA}_{\textrm{O},k} \bs \nonumber \\
  \textrm{s.t.} & \ \bs^\HH \bs = e_t, \bu^\HH \bR_\textrm{I} \bu \leq E_\textrm{I},  \bs = \bu,
\end{align}
where $\bar{\bA}_{\textrm{O},k} = -{\bA}_{\textrm{O},k}$, and $\bar{\ba}_{\textrm{O},k} = -{\ba}_{\textrm{O},k}$.
The augmented Lagrangian associated with the optimization problem in \eqref{eq:QuadSpectralADMM} can be written as
\begin{align}\label{AugLagrange}
  L_{\varrho}(\bs,\bu,\boldsymbol{\lambda})
  =& \bs^\HH \bar{\bA}_{\textrm{O},k} \bs + 2\Re(\bs^\HH \bar{\ba}_{\textrm{O},k}) + 2\Re(\boldsymbol{\lambda}^\HH(\bs-\bu)) \nonumber\\
  &+ \varrho\|\bs-\bu\|_2^2,
\end{align}
where $\boldsymbol{\lambda}$ is the Lagrange multiplier, and $\varrho$  is the penalty parameter.
The ADMM method consists of the following iterations:
\begin{subequations}\label{eq:ADMMIter}
\begin{align}
  \bs_{j+1}  &=  \arg\min_{\bs} L_{\varrho}(\bs,\bu_j,\boldsymbol{\lambda}_j), \label{eq:updateS}\\
   \bu_{j+1} &= \arg\min_{\bu} L_{\varrho}(\bs_{j+1},\bu,\boldsymbol{\lambda}_j), \label{eq:updateW}\\
    \boldsymbol{\lambda}_{j+1} &=  \boldsymbol{\lambda}_{j} + \varrho(\bs-\bu).
\end{align}
\end{subequations}

The optimization problem in \eqref{eq:updateS} can be rewritten as
\begin{align}\label{eq:ProblemUpdateS}
   \min_{\bs} &\  \bs^\HH (\bar{\bA}_{\textrm{O},k}+ \varrho \bI) \bs + 2\Re(\bs^\HH {\bb}_{\textrm{O},k}) \nonumber\\
  \textrm{s.t.} & \ \bs^\HH \bs = e_t,
\end{align}
where ${\bb}_{\textrm{O},k} = \bar{\ba}_{\textrm{O},k} + \boldsymbol{\lambda}_j - \varrho \bu_j$. The optimal solution of \eqref{eq:ProblemUpdateS} can be obtained similarly to that in \eqref{eq:QuadEnergy}:
\begin{equation}\label{eq:sj}
  \bs_{j+1} = (\bar{\bA}_{\textrm{O},k}+ (\varrho+\alpha_1) \bI)^{-1} {\bb}_{\textrm{O},k},
\end{equation}
where $\alpha_1$ is a scalar making $\bs_{j+1}^\dagger\bs_{j+1} = e_t$.

The optimization problem in \eqref{eq:updateW} is equivalent to
\begin{align}\label{eq:ProblemUpdateW}
   \min_{\bu} &\  \| \bu - \bd_j \|_2^2 \nonumber \\
  \textrm{s.t.} & \ \bu^\HH \bR_\textrm{I} \bu \leq E_\textrm{I},
\end{align}
where $\bd_j = \bs_{j+1} +\boldsymbol{\lambda}_{j}/ \varrho$.
Its solution is given by (see \cite{my2018admm} for more details):
\begin{equation}\label{eq:uj}
  \bu_{j+1} =
  \begin{cases}
    \bd_j, & \textrm{if} \ \bd_j^\dagger \bR_\textrm{I} \bd_j \leq E_\textrm{I}, \\
    (\bI + \alpha_2 \bR_\textrm{I})^{-1}\bd_j, & \textrm{if} \ \bd_j^\dagger \bR_\textrm{I} \bd_j > E_\textrm{I},
  \end{cases}
\end{equation}
where $\alpha_2$ can be obtained by solving $\bd_j^\dagger  \bT _\textrm{I} \bd_j = E_\textrm{I}$, and $\bT _\textrm{I} = (\bI + \alpha_2 \bR_\textrm{I})^{-1}\bR_\textrm{I}(\bI + \alpha_2 \bR_\textrm{I})^{-1}$.

Algorithm \ref{Alg:Spectral} summarizes the proposed ADMM algorithm for \eqref{eq:QuadSpectral}, where we terminate the ADMM algorithm if both the (Euclidian) norm of the primal residual $\br_{j}$ and the dual residual $\bd_{j}$  are sufficiently small, where $\br_{j} = \bs_j - \bu_j$, and $\bd_{j} = \varrho (\bs_j - \bs_{j+1})$.

\begin{algorithm}[!htp]
  \caption{ \small  Optimization algorithm for \eqref{eq:QuadSpectral}.}\label{Alg:Spectral}
  \KwIn{$\bA_{\textrm{O},k},\ba_{\textrm{O},k}, e_t, E_\textrm{I}$.}
  \KwOut{$\bs_{k+1}$.}
  \textbf{Initialize:} $\varrho$, $\bu = \bzero$, $\boldsymbol{\lambda} = \bzero$, $j= 0$, $\bs_{k,j} = \bs_{k}$.\\
    \Repeat{convergence}{
    Update $\bs_{k,j+1}$ by \eqref{eq:sj} \\
    $\bd_{k,j} = \bs_{k,j+1} +\boldsymbol{\lambda}_{j}/ \varrho$ \\
    Update $\bu_{k,j+1}$ by \eqref{eq:uj} \\
    $\boldsymbol{\lambda}_{j+1} =  \boldsymbol{\lambda}_{j} + \varrho(\bs_{k,j+1}-\bu_{k,j+1})$ \\
    $j=j+1$
    }
    $\bs_{k+1}$ = $\bs_{k,j}$
\end{algorithm}

\section{Algorithm Summary and Some Discussions} \label{sec:Discussion}
\subsection{Algorithm Summary and Convergence}

\begin{algorithm}[!htp]
  \caption{ \small  Waveform optimization algorithm for radar range profiling.}\label{Alg:1}
  \KwIn{$\bR_\txh,\bR_\txn$.}
  \KwOut{$\bs$.}
  \textbf{Initialize:} $k= 0$, $\bs_k$.\\
    \Repeat{convergence}{
    \Switch{Design metric}{
    \uCase{Maximizing mutual information}{
    Compute $\bG_k^{22}$ using \eqref{Gk22} and $\bG_k^{21}$ using \eqref {Gk21} \\
    Compute $\bA_{\textrm{I},k}$ and $\ba_{\textrm{I},k}$\\
    $\bA_{\textrm{O},k} = \bA_{\textrm{I},k}$, $\ba_{\textrm{O},k} = \ba_{\textrm{I},k}$
    }
    \Case{Minimizing MMSE}{
    Compute $\bH_k$ and $\bT_k$\\
    Compute $\bA_{\textrm{E},k}$ and $\ba_{\textrm{E},k}$\\
    $\bA_{\textrm{O},k} = \bA_{\textrm{E},k}$, $\ba_{\textrm{O},k} = \ba_{\textrm{E},k}$
    }
    } % end switch design metric
    \Switch{Constraint}{
    \uCase{Energy constraint}{
    Update $\bs_{k+1}$ by \eqref{eq:solutionEnergy}
    }
    \uCase{PAPR constraint}{
    Update $\bs_{k+1}$ by Algorithm \ref{Alg:PAPR}%iteratively solving \eqref{eq:QuadPAPRMinorized}
    }
    \Case{Spectral constraint}{
    Update $\bs_{k+1}$ by Algorithm \ref{Alg:Spectral}%solving \eqref{eq:QuadSpectral} with the proposed ADMM method
    }
    } % end switch constraint
    $k=k+1$
    }
\end{algorithm}
%\subsection{Convergence Analysis}
We summarize the proposed algorithm framework in Algorithm \ref{Alg:1}. Note that in some applications, to control the shape of the ambiguity function or achieve some desired property, one needs to enforce a similarity constraint on the waveform \cite{Li2006SWORD}, which can be written as
\begin{equation}\label{eq:Similarity}
  \bs^\HH \bs = e_t, \|\bs - \bs_0\|_2^2 \leq \varepsilon_2,
\end{equation}
where $\bs_0$ denotes a reference waveform possessing certain desirable property, $\bs_0^\HH \bs_0 = e_t$, and $\varepsilon_2$ is a user specified similarity parameter ($0\leq \varepsilon_2 \leq 2e_t$). One can also enforce both constant-modulus and similarity constraints on the waveform \cite{DeMaio2009Similarity,Aubry2020MultiSpectral}. The associated constraints are given by
\begin{equation}\label{eq:CMSimilarity}
  |s_l| = a_s, l=1,\cdots,L, \|\bs - \bs_0\|_\infty \leq \varepsilon_\infty,
\end{equation}
where the reference waveform $\bs_0$ is constant-modulus, and  $\varepsilon_\infty$ denotes a similarity parameter ($0\leq \varepsilon_\infty \leq 2a_s$). It can be verified that the proposed framework can be applied to deal with both the constraints in \eqref{eq:Similarity} and \eqref{eq:CMSimilarity}. Nevertheless, we omit the details here due to space limitations. %(interested readers might refer to \cite{my2018Efficient,my2019SpectralShapes} for some discussions).

Next we analyze the convergence of the proposed algorithm. Note that $f_\textrm{O}(\bs_{k+1}) \geq g_\textrm{O}(\bs_{k+1};\bs_k)$
and $f_\textrm{O}(\bs_{k}) = g_\textrm{O}(\bs_k;\bs_k)$. Thus, if $g_\textrm{O}(\bs_{k+1};\bs_k) \geq g_\textrm{O}(\bs_k;\bs_k)$, then $f_\textrm{O}(\bs_{k+1}) \geq f_\textrm{O}(\bs_{k})$, and the convergence of the sequence of the objective values is guaranteed. For the energy constraint, $g_\textrm{O}(\bs_{k+1};\bs_k) \geq g_\textrm{O}(\bs_k;\bs_k)$ owing to the optimality of $\bs_{k+1}$ given $\bs_k$; for the PAPR constraint, if $\bs_k$ is used as the starting point of the proposed MM method, the improvement of  $g_\textrm{O}(\bs_{k+1};\bs_k)$ over $g_\textrm{O}(\bs_k;\bs_k)$ can be verified by using the ascent property of the MM method; for the spectral constraint, if $\bs_{k+1}$  is obtained by the SDR and rank-one decomposition, $\bs_{k+1}$ is the optimal solution given $\bs_{k}$, and it can be checked that $g_\textrm{O}(\bs_{k+1};\bs_k) \geq g_\textrm{O}(\bs_k;\bs_k)$. Otherwise, if $\bs_{k+1}$ is obtained by the proposed ADMM method, it is non-trivial to prove the optimality of $\bs_{k+1}$. However, for the proposed ADMM algorithm, we do not encounter any convergence problem during the numerical simulations (possibly because of the hidden convexity of the problem in \eqref{eq:QuadSpectral}).
\subsection{Computational Complexity}
Table \ref{Table:1} summarizes the per-iteration computational complexity of the proposed algorithms, where $N_{papr}$ is the number of (inner) iterations needed to reach convergence under the PAPR constraint, and $N_{spectral}$ is the number of (inner) iterations needed  to reach convergence under the spectral constraint. Note that we have ignored the computational complexity that can be performed offline (e.g., the calculation of $\bR_{\textrm{n}}^{-1}$). In addition, we ignore the computational complexity involving the multiplication of $\bE$, since it can be done by addition.
%In this subsection, we analyze the per-iteration computational complexity of the proposed algorithms. If we use the mutual information as the design metric, the computation of $\bA_{\textrm{I},k}$ has a  complexity of $O(L_0^2P^2 + L_0^3)$, and the computation of $\ba_{\textrm{I},k}$ has a complexity of $O(L_0^2P+L_0P^2)$; if we use the MMSE as the design metric, the computation of $\bA_{\textrm{E},k}$ has a  complexity of $O(L_0^3 + L_0^2P^2 )$, and the computation of $\ba_{\textrm{E},k}$ has a complexity of $O(L_0^2P+L_0P^2)$. For the energy-constrained optimization problem in \eqref{eq:QuadEnergy}, the associated computational complexity is $O(L^3)$; for the PAPR-constrained optimization problem in \eqref{eq:QuadPAPR}, the method based on the MM has a complexity of $O(N_{papr}L^2)$, where $N_{papr}$ is the number of iterations needed to reach convergence for the MM method; for the spectrally constrained optimization problem in \eqref{eq:QuadSpectral}, the method based on the ADMM has a complexity of $O(N_{spectral}L^3)$, where $N_{spectral}$ is the number of iterations needed  to reach convergence for the ADMM method.

\begin{table*}[!t]
\renewcommand{\arraystretch}{1.3}
\caption{Computational complexity analysis}
\label{Table:1}
\centering
\begin{tabular}{c|c|c|c}
  \hline
  Computation& Complexity&Computation& Complexity\\
  \hline
  \multicolumn{2}{c|}{\textit{\textsf{Mutual information maximization}}} &\multicolumn{2}{c}{\textit{\textsf{MMSE minimization}}}\\
  \hline
  $\bG_k^{22}$& $O(L_0^3)$ &$\bH_k$&$O(L_0^3+L_0^2P+L_0P^2)$\\
  \hline
  $\bG_k^{21}$& $O(L_0^2P+L_0P^2)$ &$\bT_k$&$O(L_0^2P^2)$\\
  \hline
  $\bA_{\textrm{I},k}$&$O(L_0^2P^2)$ &$\bA_{\textrm{E},k}$&$O(L_0^2P^2)$\\
  \hline
  $\ba_{\textrm{I},k}$& $O(L_0^2P^2)$ &$\ba_{\textrm{E},k}$&-\\
  \hline
  \multicolumn{4}{c}{\textit{\textsf{Solving the quadratic programming problem}}}\\
  \hline
  Energy constraint& \multicolumn{3}{c}{$O(L^3)$} \\
  \hline
  PAPR constraint& \multicolumn{3}{c}{$O(N_{papr}L^2)$} \\
  \hline
  Spectral constraint& \multicolumn{3}{c}{$O(N_{spectral}L^3)$} \\
  \hline
\end{tabular}
\end{table*}
\subsection{Connection with ZCZ Waveforms}
Suppose that $\bR_\txn = \sigma_0^2\bI$ (e.g., the disturbance is dominated by the white noise) and $\bR_\txh = \boldsymbol{\Lambda}_\txh$ (i.e., the uncertainty of each element of the target impulse response vector is independent). Then the waveform design problem based on maximizing mutual information can be rewritten as
\begin{align} \label{eq:MIZCZ}
  \max_{\bs} & \ f_{{\textrm{I}}}(\bs) = \log \det (\sigma_0^{-2}\bS^\HH\bS + \boldsymbol{\Lambda}_\txh^{-1}) \nonumber \\
  \textrm{s.t.} & \ \bs \in \mathcal{S}.
\end{align}
%where $\sigma_1^2 = \sigma_h^2\sigma_0^{-2}$.
Note that
\begin{equation}%\label{}
  \bS^\HH\bS = e_t
  \begin{pmatrix}
1   & r_1^* & \cdots & r^*_{P-1} \\
r_1& 1        &\cdots  & r^*_{P-2} \\
\vdots&\vdots & \ddots & \vdots \\
r_{P-1} & r_{P-2} & \cdots & 1
  \end{pmatrix},
\end{equation}
where
$r_p = \frac{1}{e_t} \sum_{k=1}^{L-p}s_k^*s_{k+p}$
is the aperiodic correlation function of $\bs$. In addition, according to Hadamard's inequality \cite{HornJohnson1990matrixbook},
\begin{equation}%\label{}
 \det (\sigma_0^{-2}\bS^\HH\bS + \boldsymbol{\Lambda}_\txh^{-1}) \leq \prod_{p=1}^P (\sigma_0^{-2}e_t + \lambda_{\txh,p}^{-1}),
\end{equation}
where $\lambda_{\txh,p}$ is the $p$th diagonal element of $ \boldsymbol{\Lambda}_\txh$, and the equality holds if and only if $\bS^\HH\bS$ is diagonal. Thus, to maximize the mutual information, the correlation function of $\bs$ should satisfy that $r_p \approx 0, p=1, \cdots, {P-1}$. The corresponding waveform is called zero-correlation zone (ZCZ) waveform \cite{fan1999ZCZ}.

Under the same assumption,  the waveform design problem based on minimizing MMSE can be reformulated by
\begin{align} \label{eq:MMSEZCZ}
  \min_{\bs} & \ \tr\left(\left[\boldsymbol{\Lambda}_\txh^{-1} + \sigma_0^{-2}\bS^\HH\bS\right]^{-1}\right) \nonumber \\
  \textrm{s.t.} & \ \bs \in \mathcal{S}.
\end{align}
Note that \cite{Yang2007MMSE}
\begin{equation}%\label{}
  \tr\left(\left[\boldsymbol{\Lambda}_\txh^{-1} + \sigma_0^{-2}\bS^\HH\bS\right]^{-1}\right) \geq \sum_{p=1}^P \frac{1}{\lambda_{\txh,p}^{-1} + \sigma_0^{-2}e_t},
\end{equation}
where the equality holds if and only if $\bS^\HH\bS$ is diagonal. Thus, in this situation, the ZCZ waveform also minimizes the MMSE.

\subsection{Mutual Information, MMSE, and SNR}
If $P=1$ (i.e., a point-like target), the mutual information in \eqref{eq:MI} can be written as
\begin{align}%\label{}
  \log \det (\bR_\txn^{-1}\bS  \bR_\txh \bS^\HH + \bI)
  &= \log (1+ \lambda_\txh\bs^\HH \bR_\txn^{-1} \bs) \nonumber \\
  &= \log(1+\lambda_\txh \textsf{SNR}),
\end{align}
where we define $\textsf{SNR} = \bs^\HH \bR_\txn^{-1} \bs$.
On the other hand, note that the MMSE for this case is given by
\begin{align}%\label{}
  \textsf{MMSE}
  &= \frac{1}{\lambda_\txh^{-1} + \bs^\HH \bR_\txn^{-1} \bs} %\nonumber \\
  =  \frac{1}{\lambda_\txh^{-1} + \textsf{SNR} }.
\end{align}
Thus, the minimization of MMSE, the maximization of mutual information, and the maximization of SNR are equivalent for the case of $P=1$.

For the more general case that $P>1$ and $\boldsymbol{\mu}_\txh = \bzero$, if $\lambda_{\max}(\bR_\txn^{-1}\bS  \bR_\txh \bS^\HH) \ll 1$ (corresponding to the case that SNR is low), where $\lambda_{\max}(\bR_\txn^{-1}\bS  \bR_\txh \bS^\HH)$ is the largest eigenvalue of $\bR_\txn^{-1}\bS  \bR_\txh \bS^\HH$, the mutual information in \eqref{eq:MI} can be approximated by
\begin{align}%\label{}
  \log \det (\bR_\txn^{-1}\bS  \bR_\txh \bS^\HH + \bI)
&\approx \tr(\bR_\txn^{-1}\bS  \bR_\txh \bS^\HH) \nonumber \\
&= \bs^\HH \bE^\HH (\bR_\txh^* \otimes \bR_\txn^{-1} ) \bE  \bs.
\end{align}
By using the low SNR assumption, it can be checked that we can obtain the globally optimal solutions to both the energy-constrained and the spectrally constrained waveform design problems.

\section{Numerical Examples} \label{sec:Example}
In this section, we provide several numerical examples to demonstrate the performance of the proposed algorithms. Unless otherwise stated, the code length of the waveform is $L=100$. The target occupies $P=10$ range bins. The mean of the target impulse response is $\boldsymbol{\mu}_\txh = [5,5,\cdots,5]^T$ and the covariance matrix is $\bI_P$. We model the disturbance covariance matrix like that in \cite{Stoica2012joint,Soltanalian2013crew}:
\begin{equation}\label{eq:disturbance}
  \bR_\txn  = \sigma^2_\textrm{J} \bR_\textrm{J} + \sigma^2 \bI,
\end{equation}
where $\sigma^2_\textrm{J} = 1000$ and $\sigma^2 = 1$ are the jamming and noise powers, respectively, the $(m,n)$th element of the jamming covariance matrix $\bR_\textrm{J}$ is given by $q_{m-n}$ ($m\geq n$), $[q_0, q_1, \cdots, q_{L_0-1}, q_{L_0-1}^*, \cdots, q_1^*]^T$ can be obtained by the inverse discrete Fourier transform (IDFT) of $\{\eta_p\}_{p=1}^{2L_0-1}$, and $\{\eta_p\}$ denotes the jamming power spectrum at the frequency $(p-1)/(2L_0-1)$, ${p=1}, 2, \cdots, {2L_0-1}$. For simplicity, we consider a barrage jamming whose power spectrum is given by
\begin{equation}\label{eq:barragePower}
  \eta_p = \begin{cases}
    1 & \lfloor(2L_0-1)f_{1,\textrm{J}}\rfloor \leq p \leq \lfloor(2L_0-1)f_{2,\textrm{J}}\rfloor, \\
    0 & \textrm{otherwise},
  \end{cases}
\end{equation}
where $f_{1,\textrm{J}} = 0.1$, and $f_{1,\textrm{J}} = 0.3$. The available transmit energy is $e_t=L$. We terminate the proposed algorithms if $|f_\textrm{O}(\bs_{k+1})-f_\textrm{O}(\bs_{k})|/f_\textrm{O}(\bs_{k+1}) \leq \varepsilon$, where $\varepsilon = 10^{-4}$. Finally, all the analysis is carried out on a standard laptop with Intel Core i7-8550U and 8 GB RAM.

\subsection{Constant-modulus Constraint}
\begin{figure}[!htp]
\centering
{\subfigure[]{{\includegraphics[width = 0.35\textwidth]{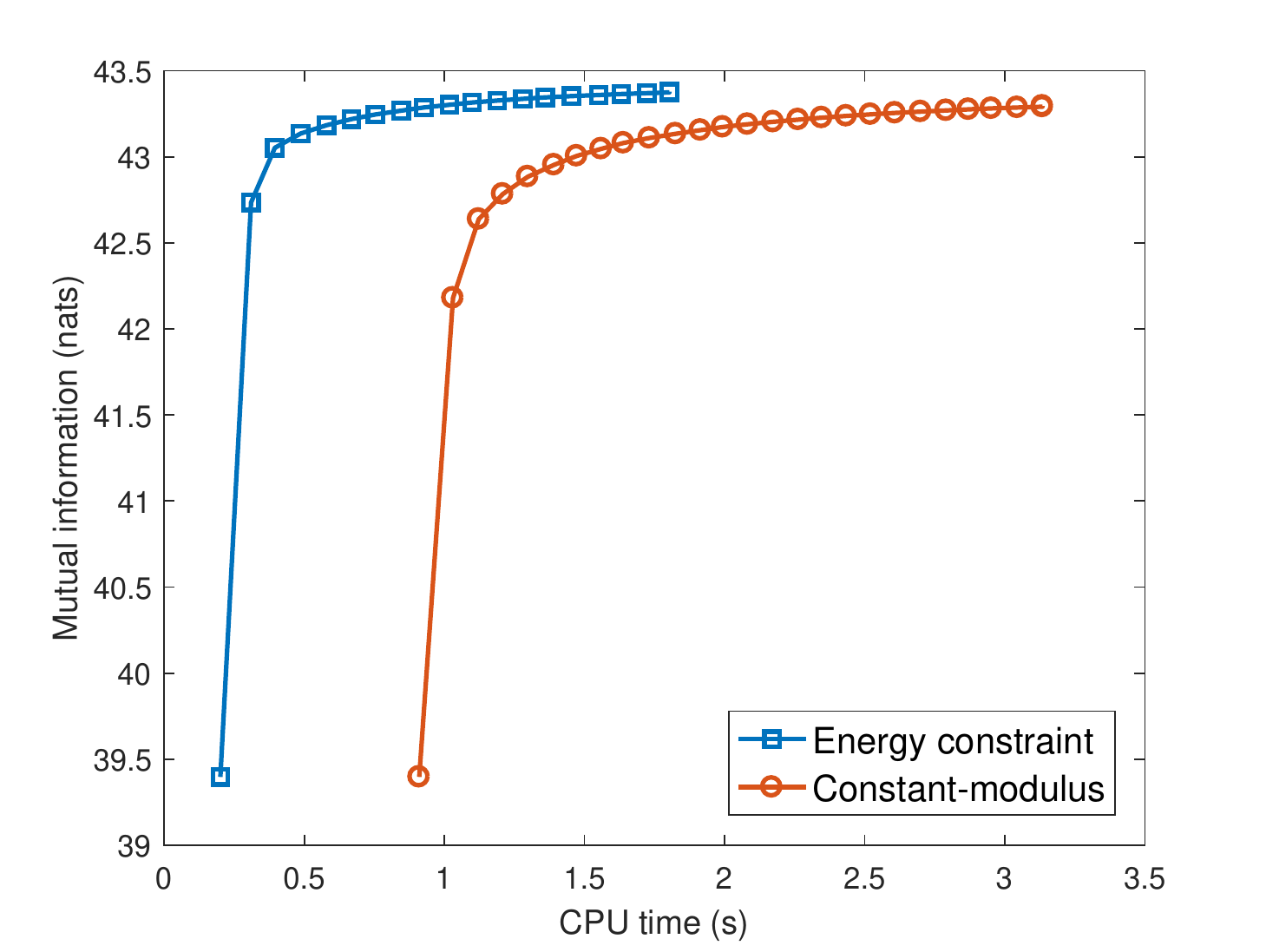}} \label{Fig:1a}} }
{\subfigure[]{{\includegraphics[width = 0.35\textwidth]{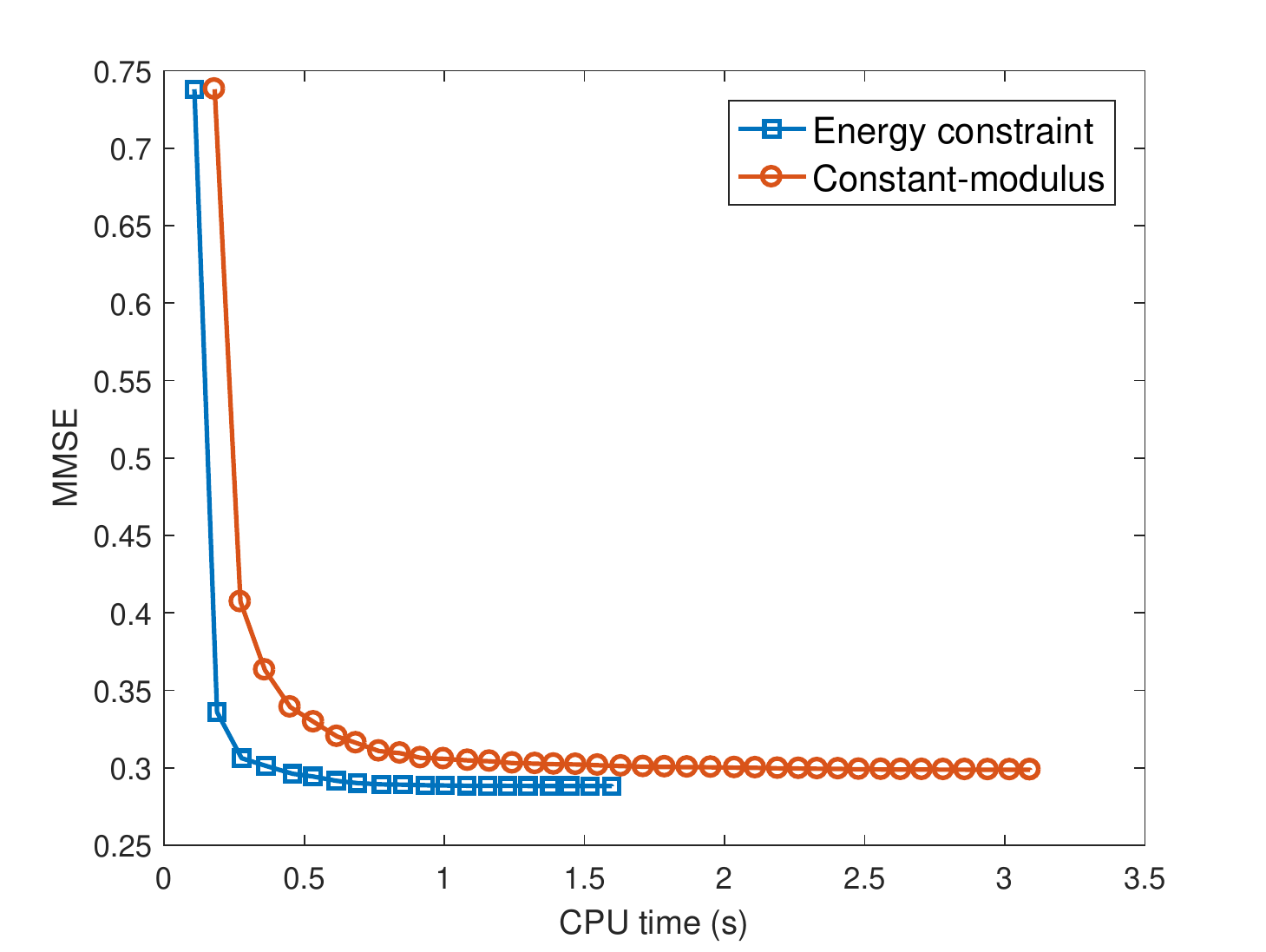}} \label{Fig:1b}} }
\caption{Convergence of the objective values versus CPU time. PAPR constraint. $\rho=1$. (a) Mutual information. (b) MMSE.}
\label{Fig:1}
\end{figure}

\begin{figure}[!htp]
\centering
{\subfigure[]{{\includegraphics[width = 0.35\textwidth]{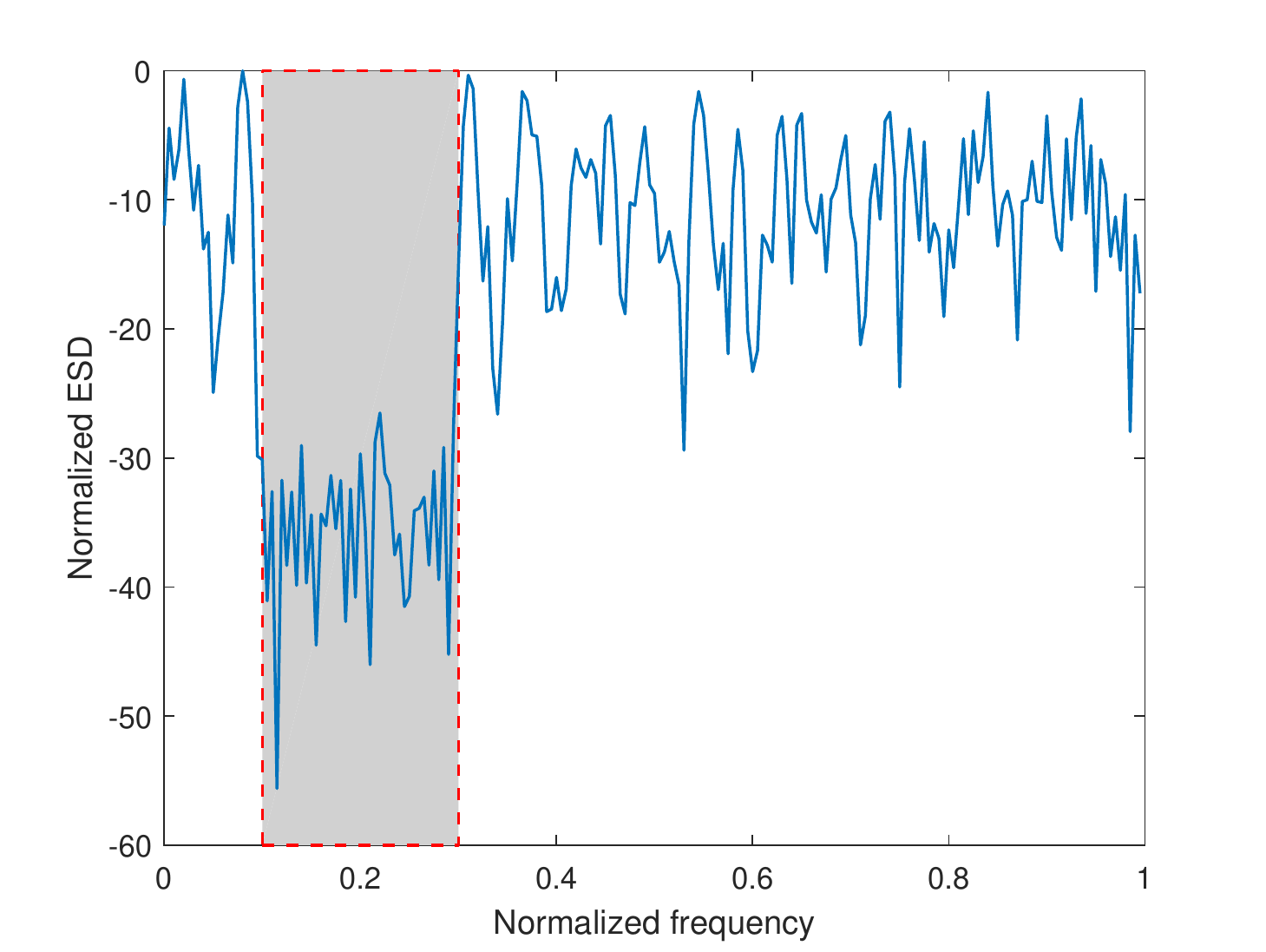}} \label{Fig:2a}} }
{\subfigure[]{{\includegraphics[width = 0.35\textwidth]{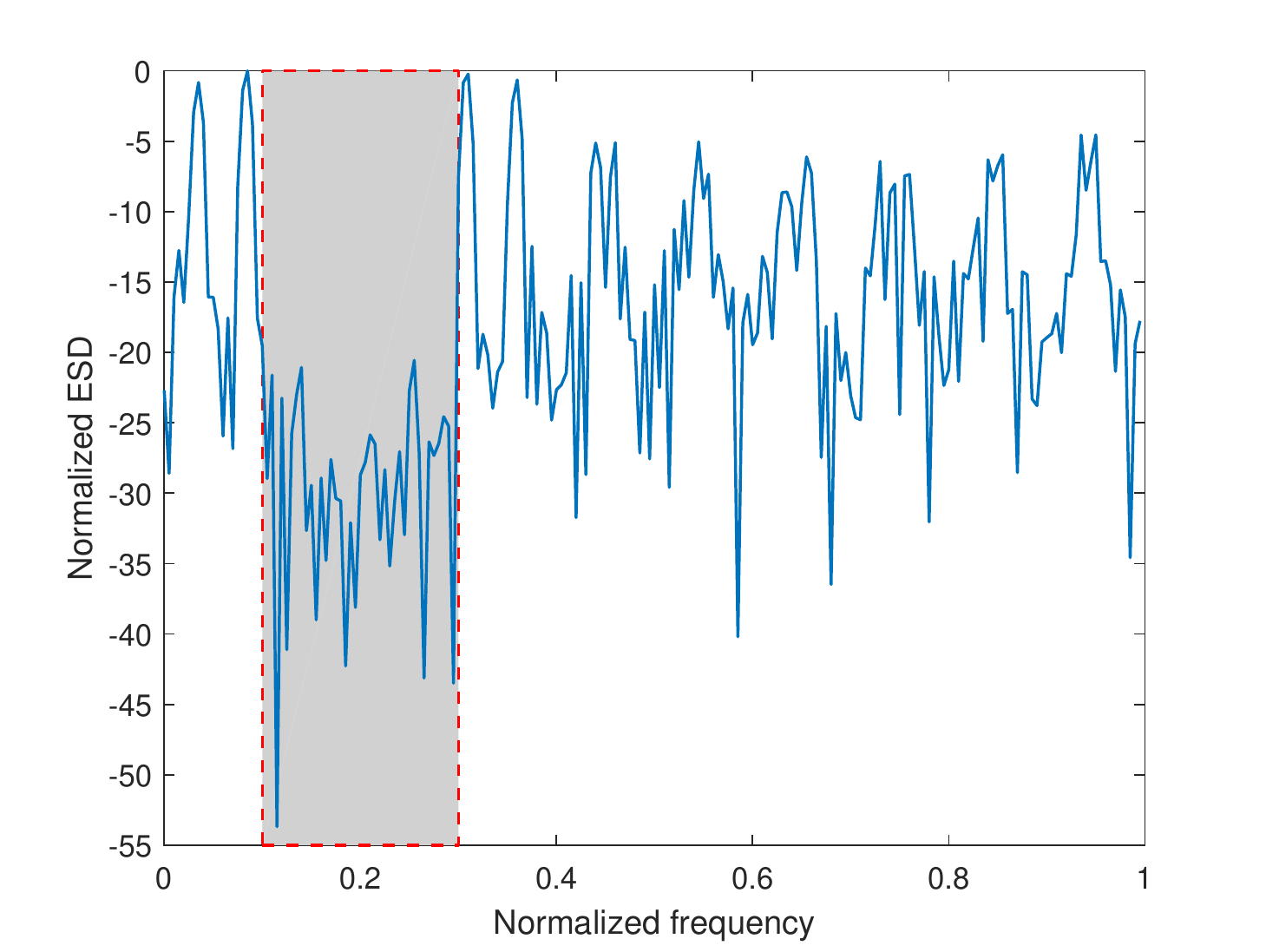}} \label{Fig:2b}} }
\caption{The ESD of the synthesized constant-modulus waveform at convergence. (a) Mutual information. (b) MMSE.}
\label{Fig:2}
\end{figure}

In this subsection, we analyze the performance of the synthesized constant-modulus waveforms. Fig. \ref{Fig:1} shows the objective values (including mutual information in Fig. \ref{Fig:1a} and MMSE in Fig. \ref{Fig:1b}) of the waveforms synthesized by the proposed algorithms versus CPU time, where we initialize the proposed algorithms with the LFM waveform (i.e., the $k$th element of the waveform is given by $\sqrt{e_t/L} \exp(j\pi(k-1)^2/L)$). Moreover, we plot the objective values associated with the energy-constrained waveforms as a benchmark. Note that the curves associated with the objective values (mutual information or MMSE) have monotonically (increasing/decreasing) behaviors. In addition, the objective values  of the synthesized constant-modulus waveforms at convergence are close to those of the energy-constrained waveforms. In Fig. \ref{Fig:2}, we plot the (normalized) energy spectral density (ESD) of the constant-modulus waveforms synthesized by the proposed algorithms. Interestingly, we can observe that the ESDs of the synthesized waveforms form a notch in the frequency band where the barrage jamming exists (shaded in light gray). However, the relationship of the notch depth in this frequency band with the jamming power is unknown.

\begin{figure}[!htp]
\centering
{\subfigure[]{{\includegraphics[width = 0.35\textwidth]{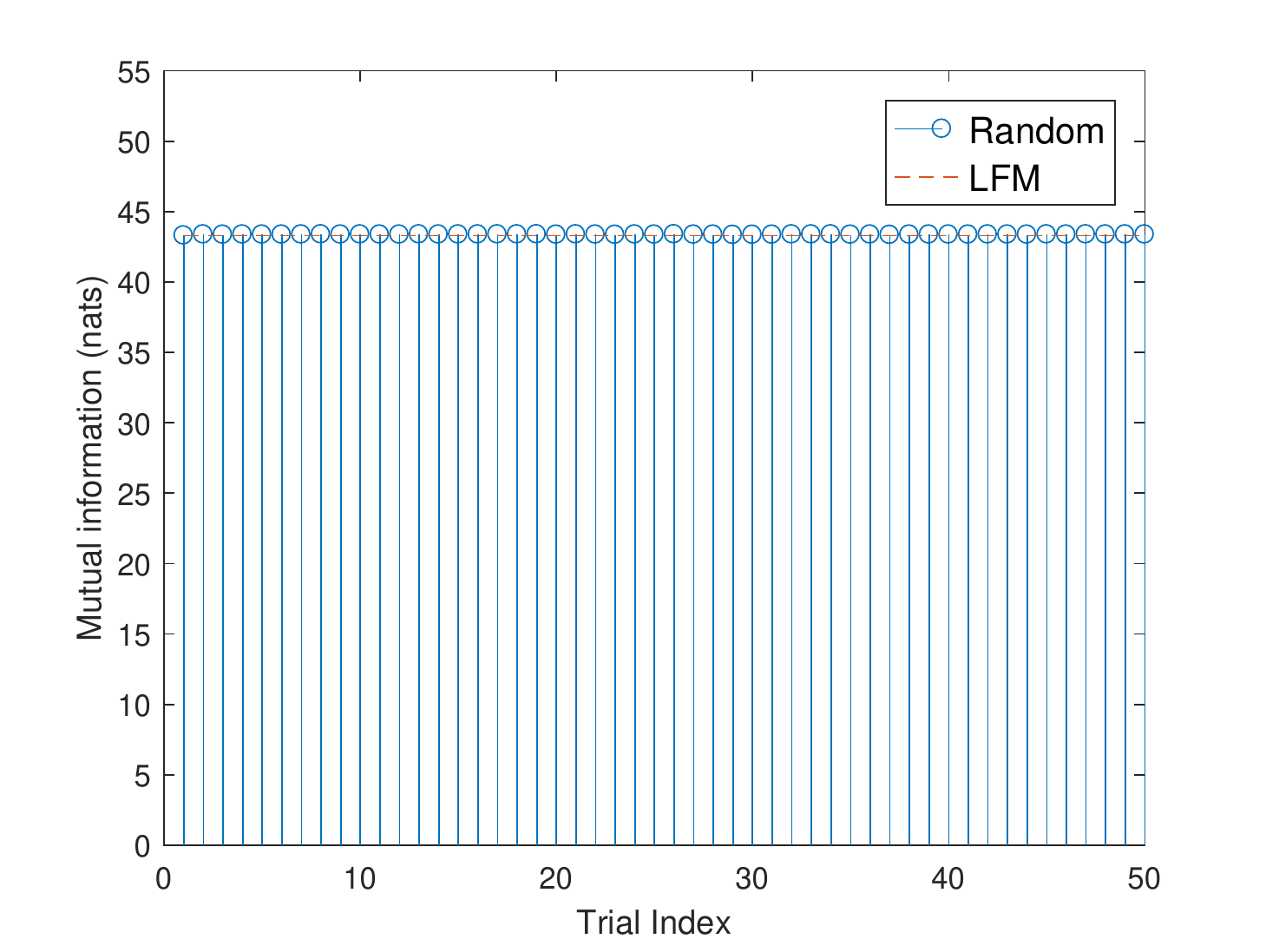}} \label{Fig:3.5a}} }
{\subfigure[]{{\includegraphics[width = 0.35\textwidth]{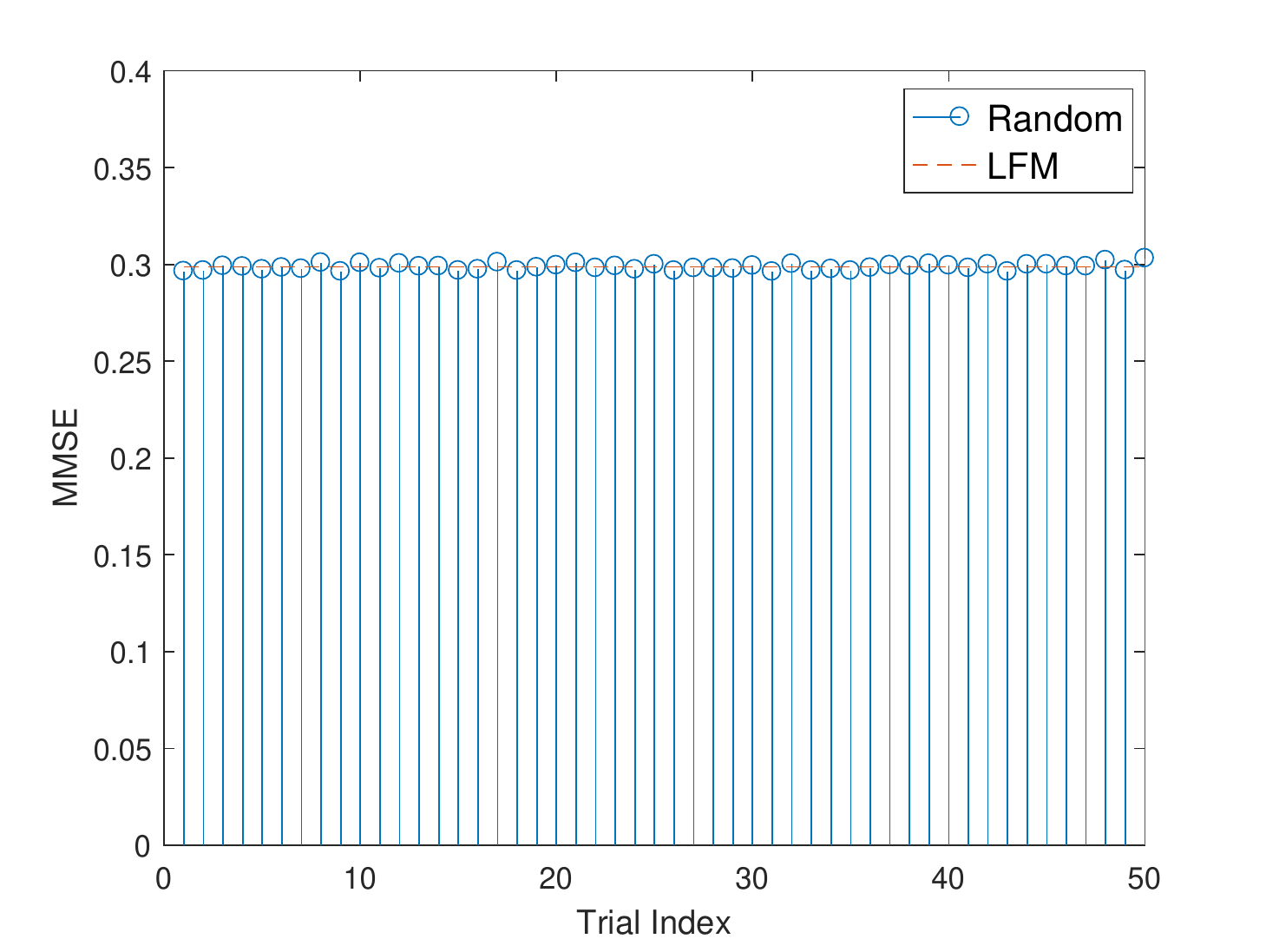}} \label{Fig:3.5b}} }
\caption{ The impact of starting points on the performance of the synthesized waveforms. (a) Mutual information. (b) MMSE.}
\label{Fig:2.5}
\end{figure}

To analyze the impact of starting points on the performance of the proposed algorithms, we use the same parameters as Fig. \ref{Fig:1} and conduct 50 independent Monte Carlo runs using random-phase waveforms as starting points for $\bs$. Specifically, the magnitude of the random-phase waveforms is $a_s$, and the phases are independent random variables uniformly distributed in $[0, 2\pi]$. Fig. \ref{Fig:2.5} shows the objective values  (mutual information or MMSE) at convergence for different starting points, and plots the values of the synthesized waveforms in Fig. \ref{Fig:1} (which is initialized by LFM). We notice that for all the runs, the objective values at convergence are almost identical to that of the synthesized waveforms in Fig. \ref{Fig:1}. Therefore, the proposed algorithms are insensitive to the starting points.

\begin{figure}[!htp]
\centering
\centering
{\subfigure[]{{\includegraphics[width = 0.35\textwidth]{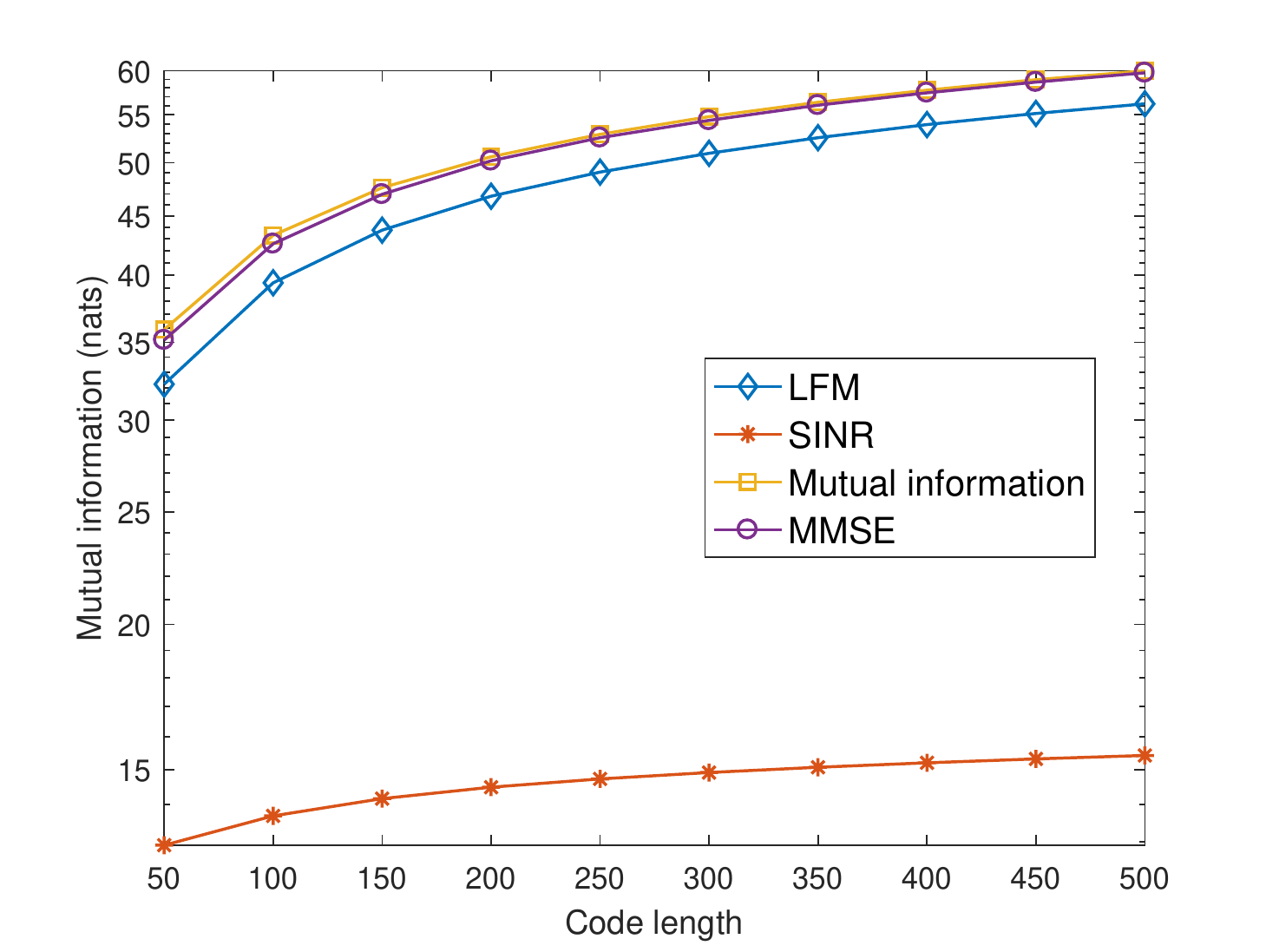}} \label{Fig:3a}} }
{\subfigure[]{{\includegraphics[width = 0.35\textwidth]{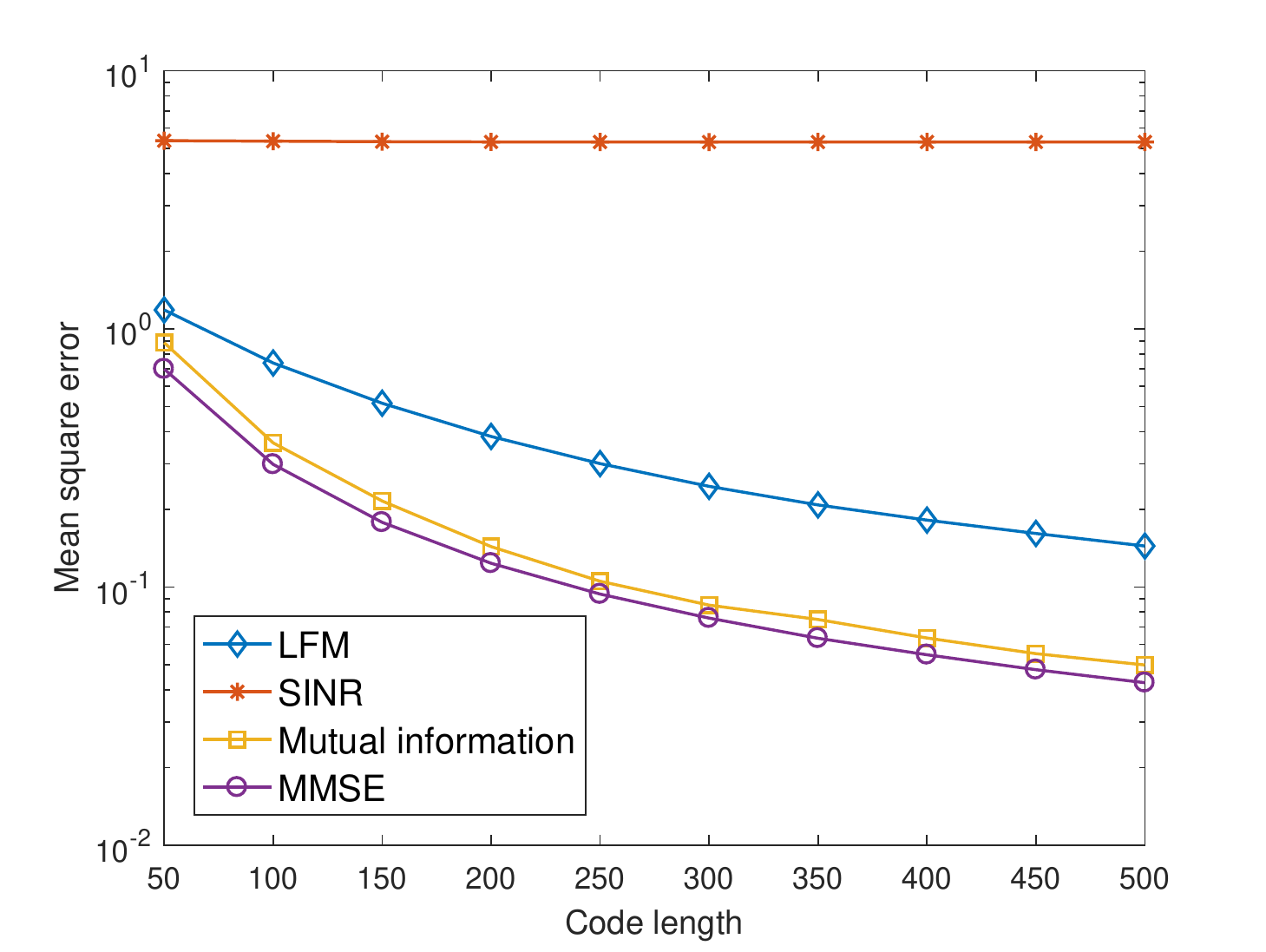}} \label{Fig:3b}} }
\caption{The performance of the synthesized constant-modulus waveforms versus code length. $e_t = L$. (a) Mutual information. (b) MMSE.}
\label{Fig:3}
\end{figure}

Next we compare the performance of the constant-modulus waveforms synthesized by the proposed algorithms with that of the LFM waveform as well as the waveform designed based on maximizing the signal to interference-plus-noise ratio (SINR) (see, e.g., \cite{Garren2002matchedillumination,Cheng2017PolarimetricDesign,Stoica2012joint,DeMaio2011PAPR} for extensive discussions on this topic). We use a similar method to that in \cite{Garren2002matchedillumination} to synthesize the waveform maximizing the SINR. Note that the algorithm in \cite{Garren2002matchedillumination} cannot deal with the constant-modulus constraint. To adapt the algorithm in \cite{Garren2002matchedillumination} to tackle the constant-modulus waveform design problem, we resort to Algorithm \ref{Alg:PAPR} proposed in Section \ref{Subsection:PAPR}. Fig. \ref{Fig:3} shows the performance of different waveforms versus the code length. For each waveform, the available transmit energy is $e_t = L$ (i.e., the transmit energy  is the same for different waveforms but varies with the code length). We can observe that the performance of the waveforms designed based on maximizing SINR is poor. Note that such waveforms are designed to enhance the detection performance of the radar systems, but not for range profiling. Thus, it is important to adapt the transmit waveform according to the radar tasks. In addition, the waveforms  based on maximizing mutual information achieve the largest mutual information and the waveforms based on minimizing the MMSE achieve the smallest MSE. Interestingly, although different criteria are used, the performance of the waveforms designed based on maximizing mutual information is close to that of the waveforms based  on minimizing MMSE. Moreover, both kinds of waveforms outperform the conventional LFM waveforms.

\subsection{Spectral Constraint}
\begin{figure}[!htp]
\centering
{\subfigure[]{{\includegraphics[width = 0.35\textwidth]{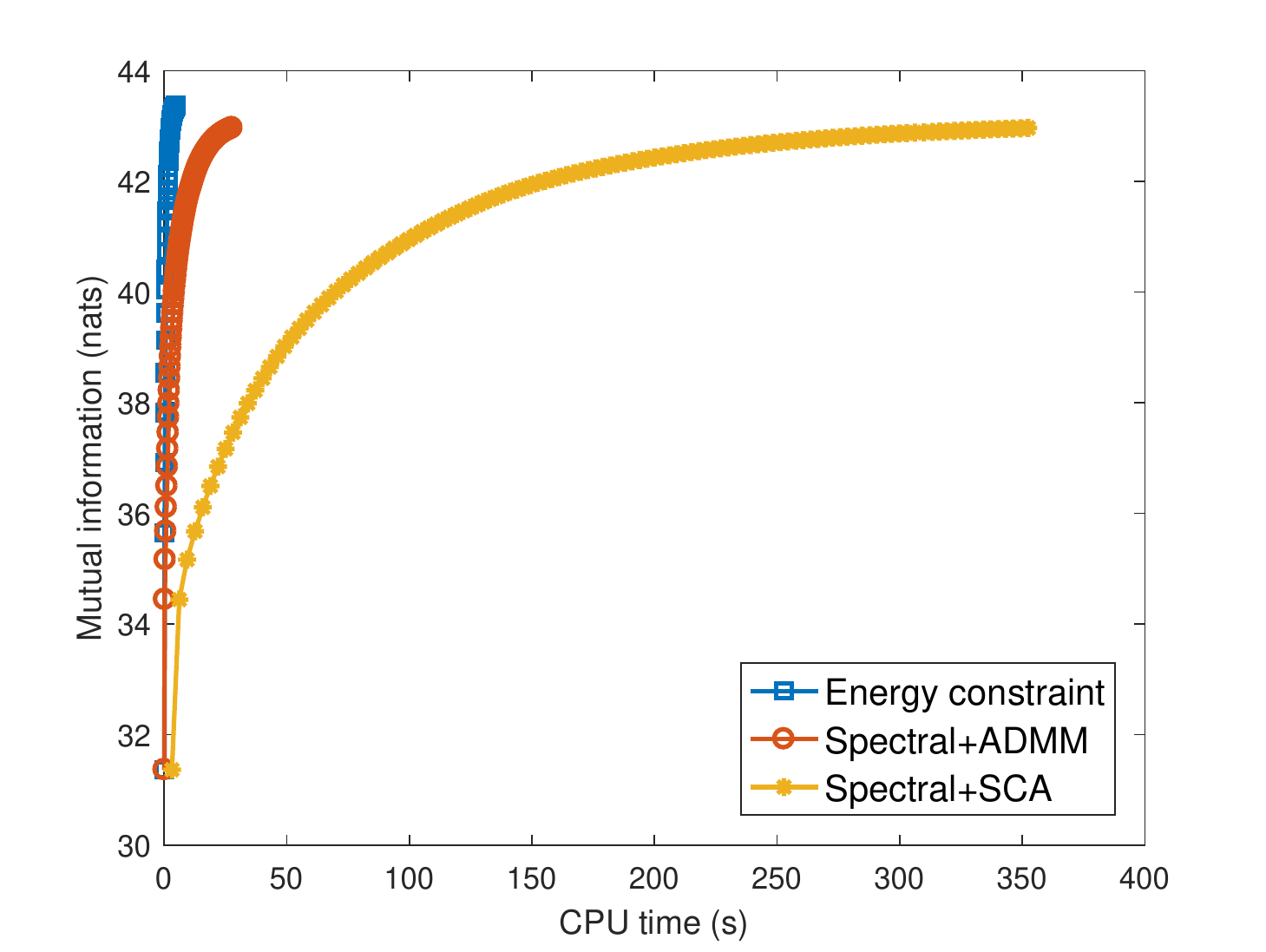}} \label{Fig:4a}} }
{\subfigure[]{{\includegraphics[width = 0.35\textwidth]{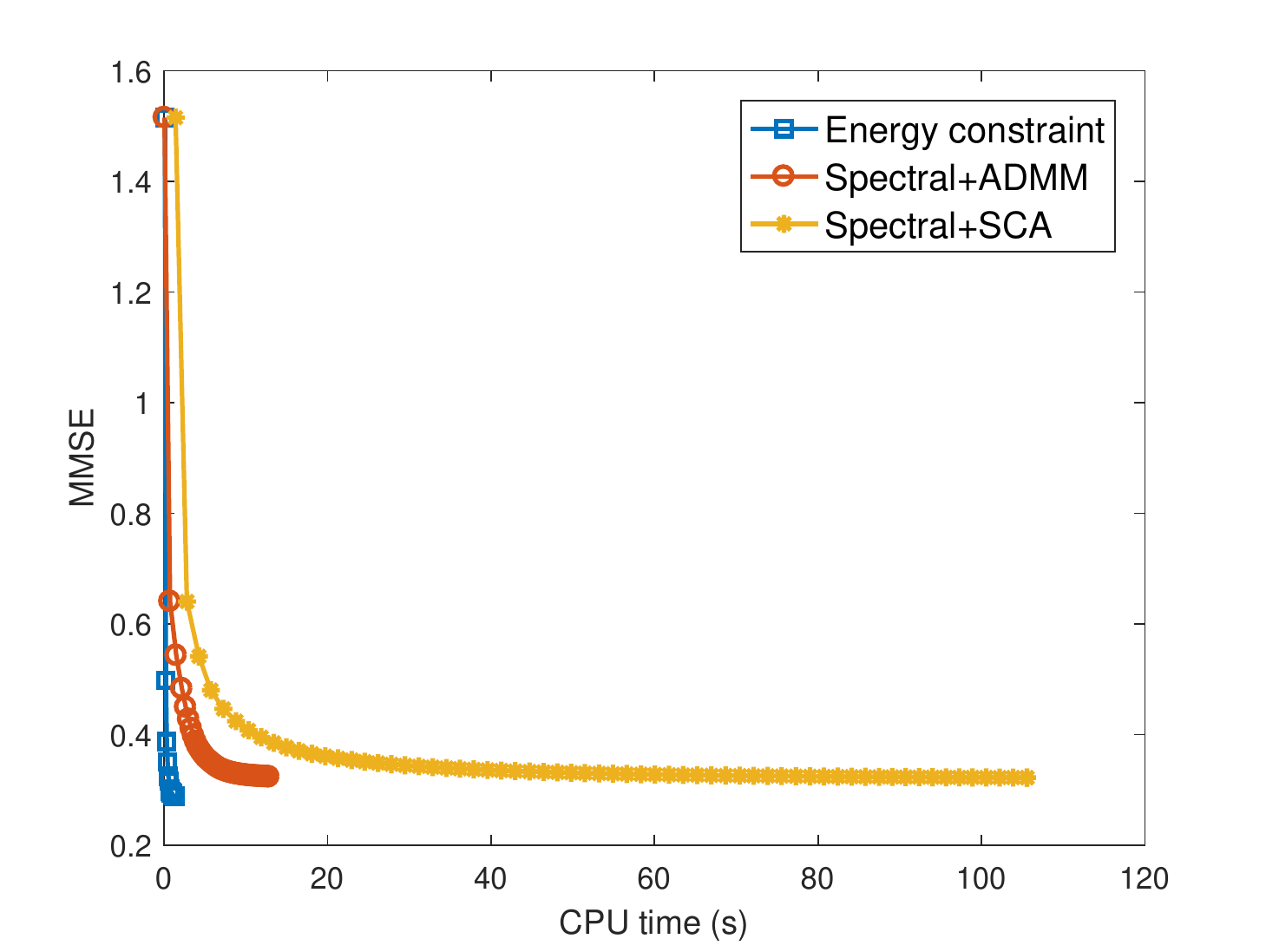}} \label{Fig:4b}} }
\caption{Convergence of the objective values versus the number of iterations. Spectral constraint. $E_\textrm{I} = 0.05$. (a) Mutual information. (b) MMSE.}
\label{Fig:4}
\end{figure}

\begin{figure}[!htp]
\centering
{\subfigure[]{{\includegraphics[width = 0.35\textwidth]{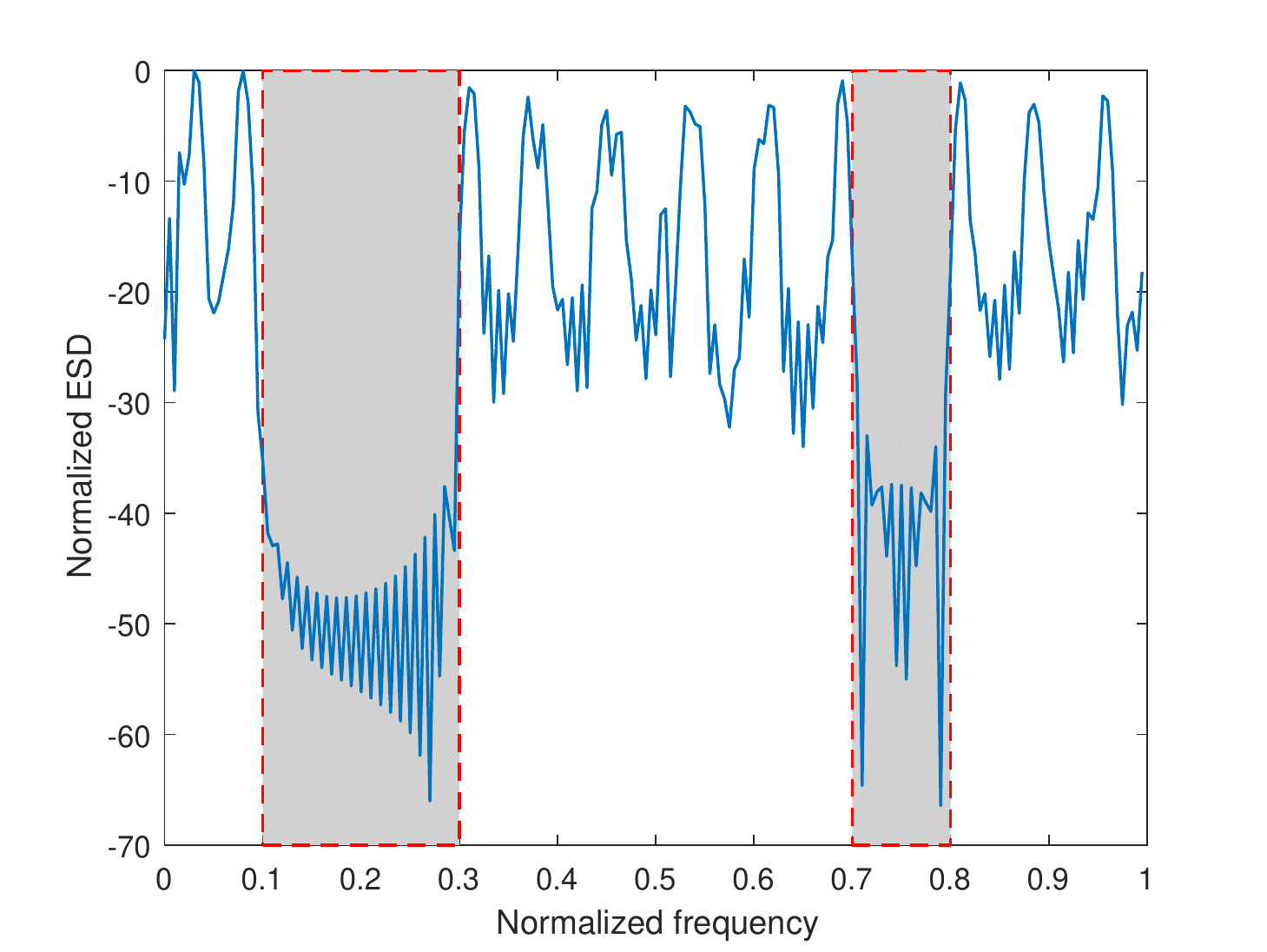}} \label{Fig:5a}} }
{\subfigure[]{{\includegraphics[width = 0.35\textwidth]{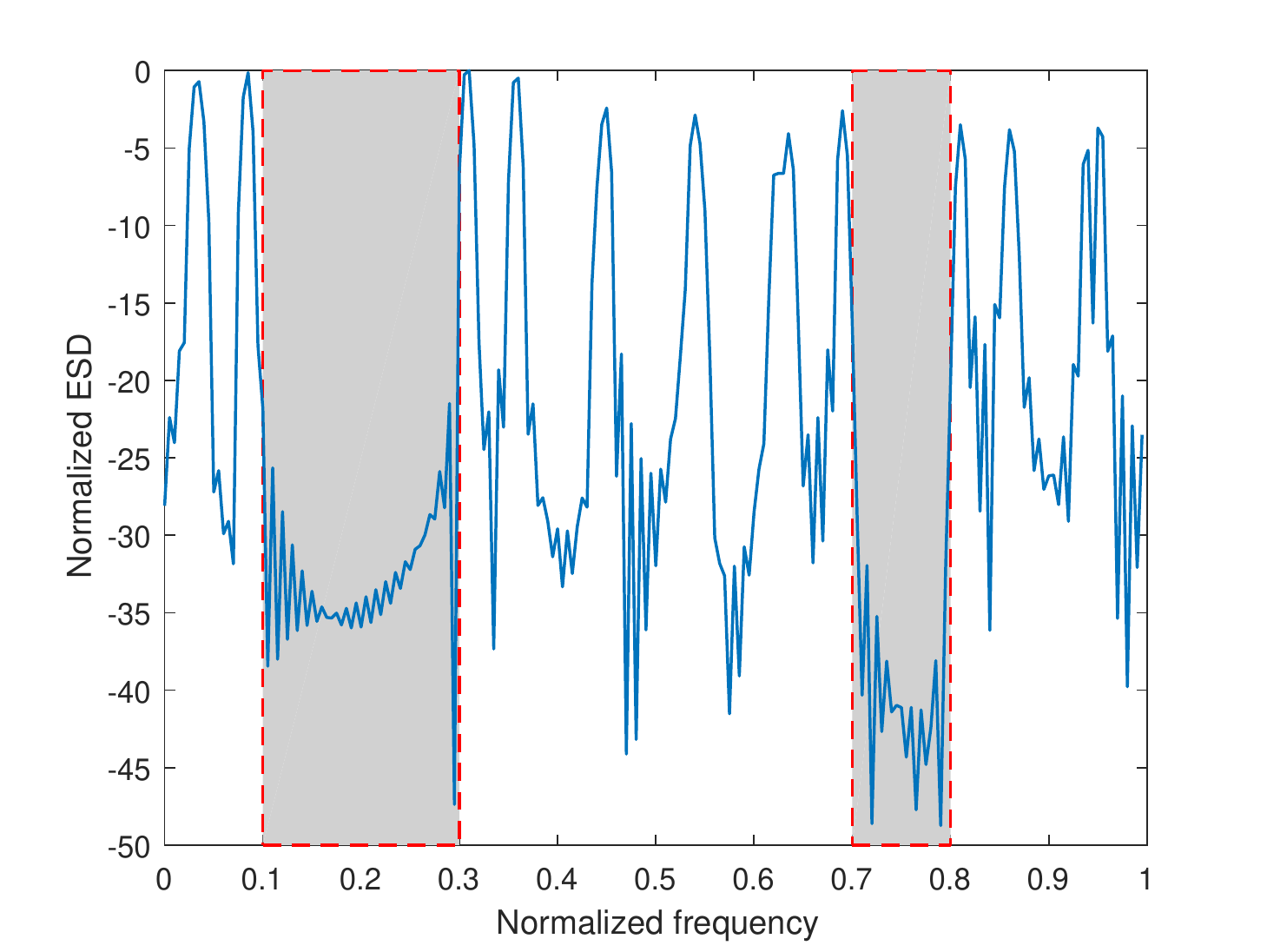}} \label{Fig:5b}} }
\caption{The ESD of the synthesized spectrally constrained waveform at convergence. $E_\textrm{I} = 0.05$. (a) Mutual information. (b) MMSE.}
\label{Fig:5}
\end{figure}

In this subsection, we analyze the performance of the spectrally constrained waveforms synthesized by the proposed algorithms. Fig. \ref{Fig:4} shows the objective values of the synthesized waveforms versus CPU time, where a licensed communication system operates in the (normalized) frequency band $[0.7, 0.8]$, the maximum allowed interference on the communication band is $E_\textrm{I} = 0.05$, and we initialize the proposed algorithms with the scaled eigenvector associated with the smallest eigenvalue of $\bR_\textrm{I}$. To tackle the spectrally constrained quadratic programming problem in \eqref{eq:QuadSpectral}, we apply the  ADMM algorithm proposed in Subsection \ref{Subsection:Spectral} and compare it with the FPP-SCA algorithm proposed in \cite{Wu2018MultipleConstraints}. In the ADMM algorithm, we set $\varrho=1$. In the FPP-SCA algorithm, the penalty parameter is 10. In addition,  we plot the objective values associated with the energy-constrained waveform as a benchmark. We can observe the convergence of the objective values for all the cases. Moreover, compared with the energy-constrained waveforms, the additional spectral constraint on the waveform results in insignificant performance loss. Furthermore, the application of the proposed ADMM algorithm results in much faster convergence than that of the FPP-SCA algorithm. This is due to the lower per-iteration computational complexity of the ADMM algorithm ($O(L^{3})$) than that of the FPP-SCA algorithm ($O(L^{3.5})$).  Fig. \ref{Fig:5} shows the ESDs of the synthesized spectral-constrained waveforms at convergence. Note the spectral notch in the communication band (shaded in light gray) and that the interference energy in this band can be controlled ($\leq E_\textrm{I} $). Thus, the waveforms synthesized by the proposed algorithms enable better compatability with the nearby communication systems.

\begin{figure}[!htp]
\centering
\centering
{\subfigure[]{{\includegraphics[width = 0.35\textwidth]{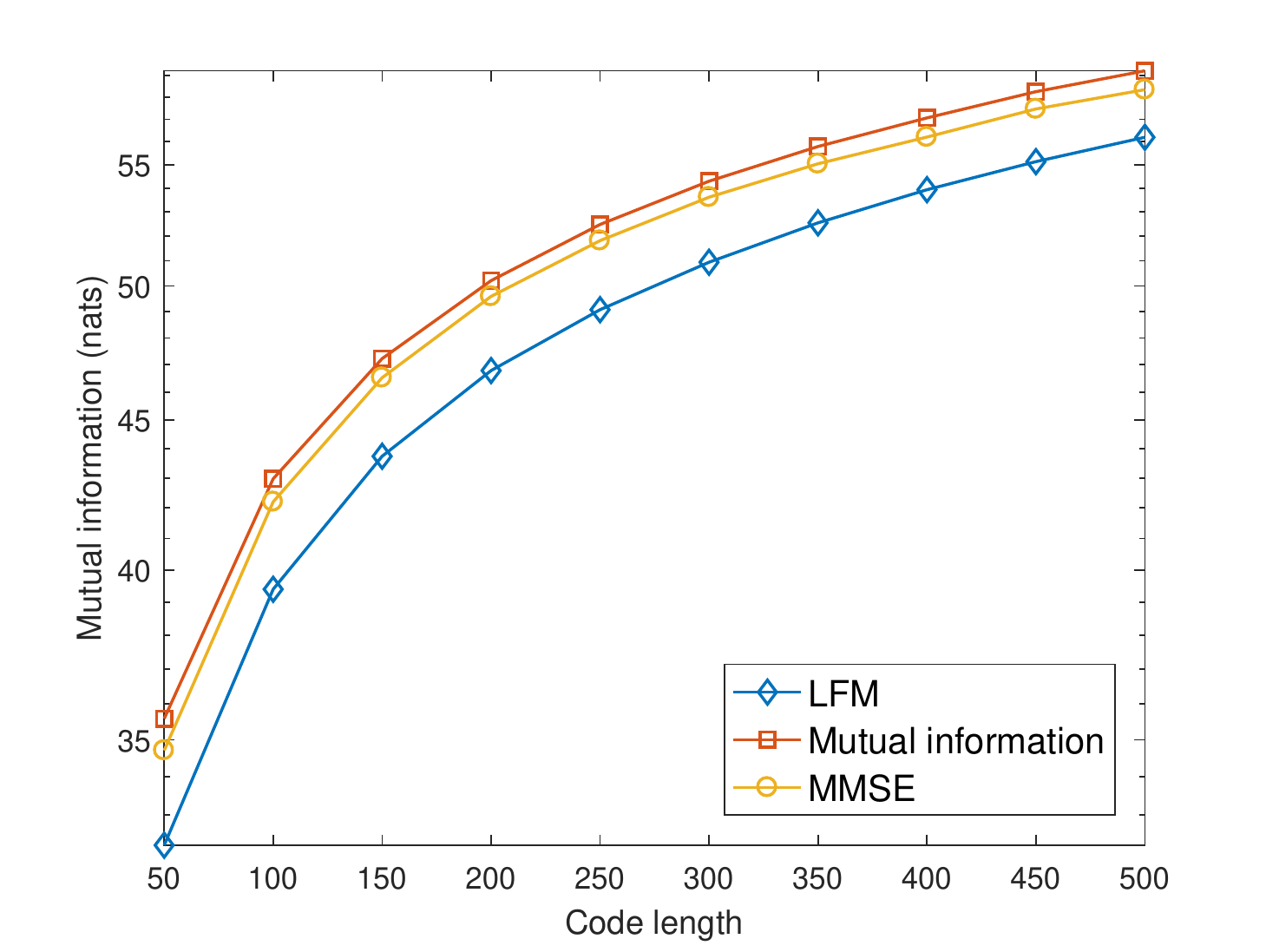}} \label{Fig:6a}} }
{\subfigure[]{{\includegraphics[width = 0.35\textwidth]{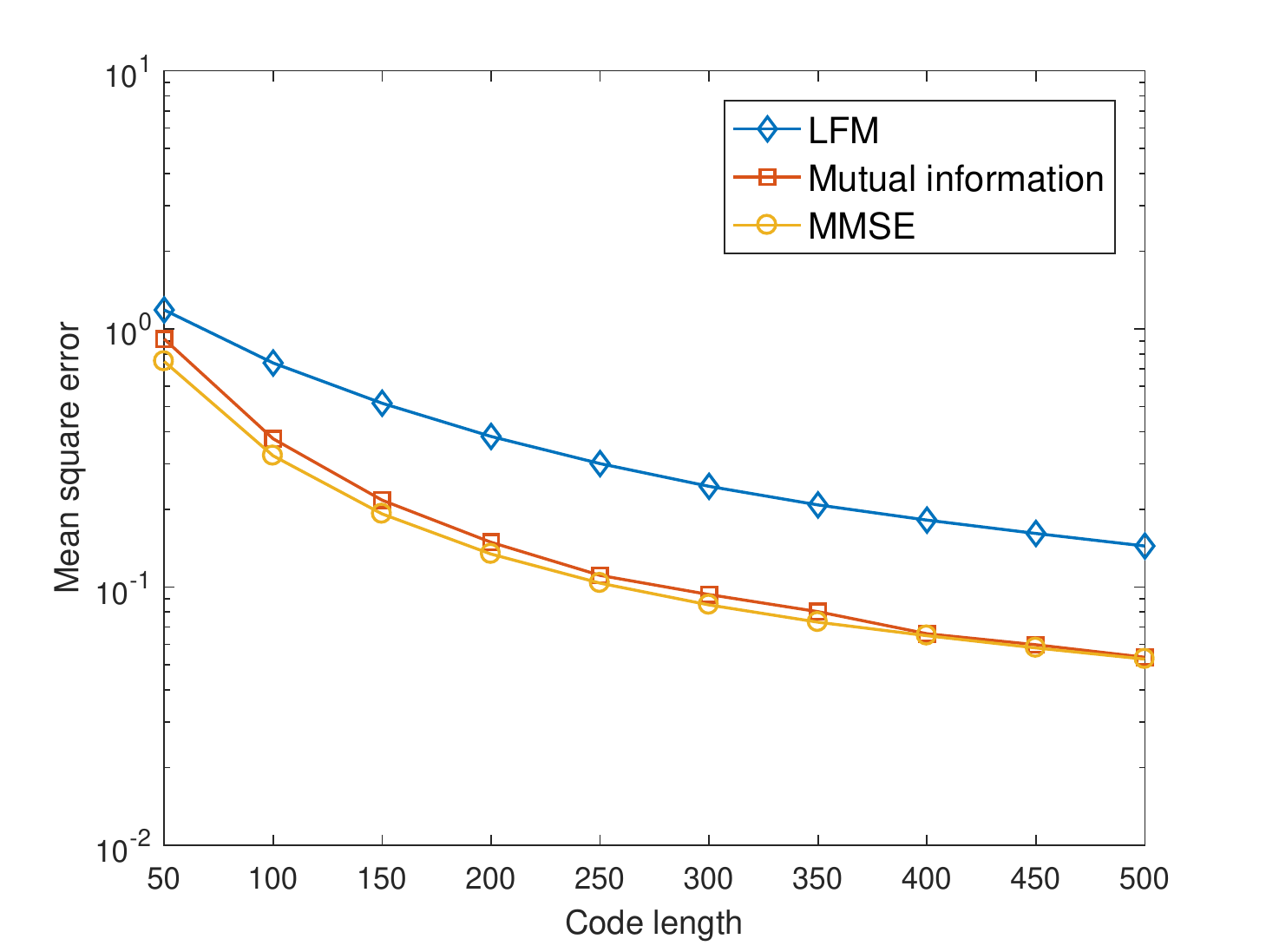}} \label{Fig:6b}} }
\caption{The performance of the synthesized spectrally constrained waveforms versus code length. $E_\textrm{I} = 0.05$. (a) Mutual information. (b) MMSE.}
\label{Fig:6}
\end{figure}

Fig. \ref{Fig:6} compared the MSE of the spectrally constrained waveforms with that of the LFM waveforms for different code lengths, where $e_t = L$, and $E_\textrm{I} = 0.05$. The results indicate that both kinds of waveforms not only have better spectral compatability than the LFM waveforms, but also achieve smaller MSE. Moreover, for large code length, the performance of the waveform based on maximizing mutual information is almost identical  to that the waveform based on minimizing MMSE.

\subsection{ZCZ Waveforms}
\begin{figure}[!htp]
\centering
{\subfigure[]{{\includegraphics[width = 0.35\textwidth]{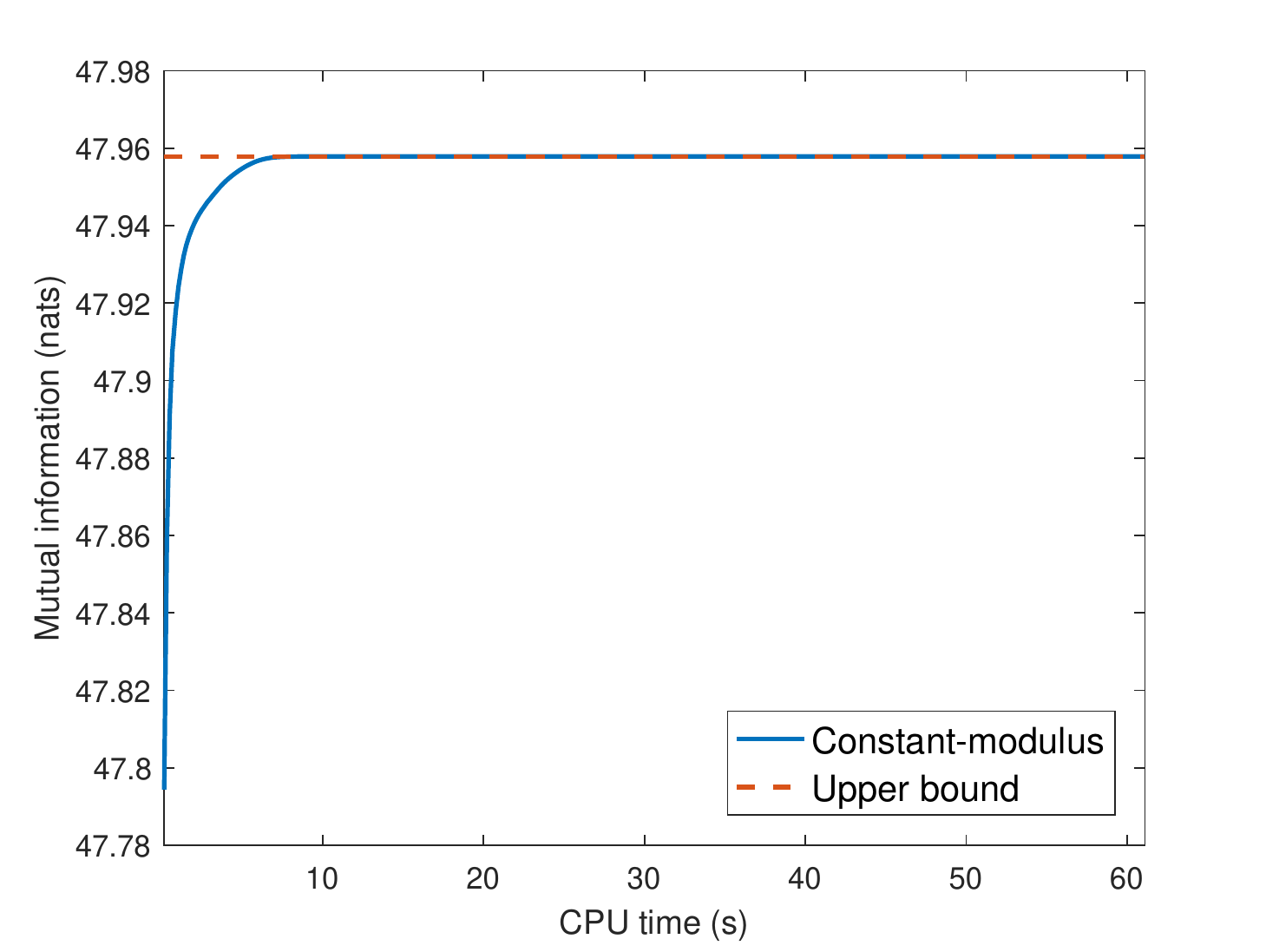}} \label{Fig:7a}} }
{\subfigure[]{{\includegraphics[width = 0.35\textwidth]{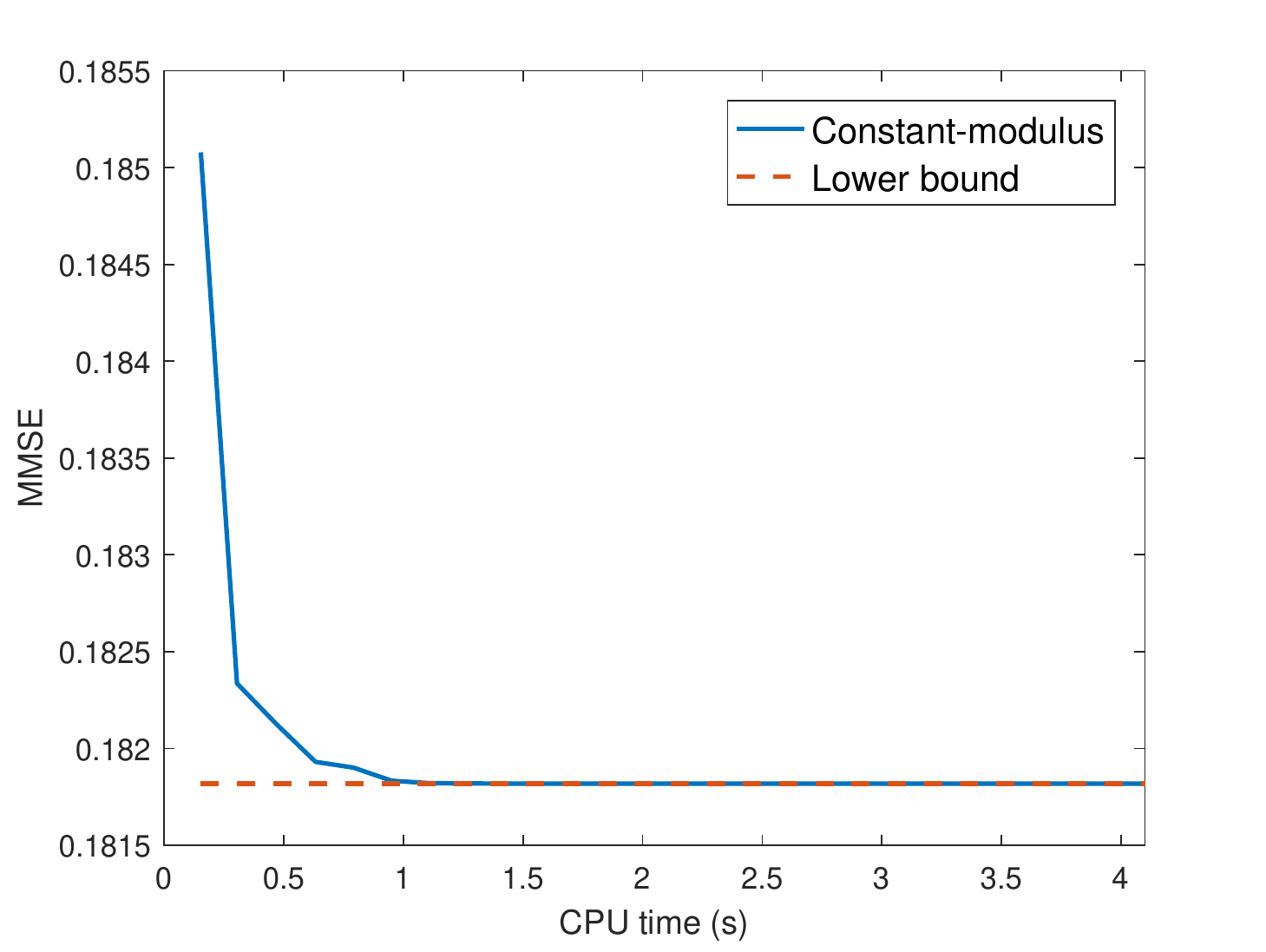}} \label{Fig:7b}} }
\caption{Convergence of the objective values versus the number of iterations. $\rho = 1$. $\bR_\txh = 0.1 \bI$. $\bR_\txn = \bI$. (a) Mutual information. (b) MMSE.}
\label{Fig:7}
\end{figure}

\begin{figure}[!htp]
\centering
{\subfigure[]{{\includegraphics[width = 0.35\textwidth]{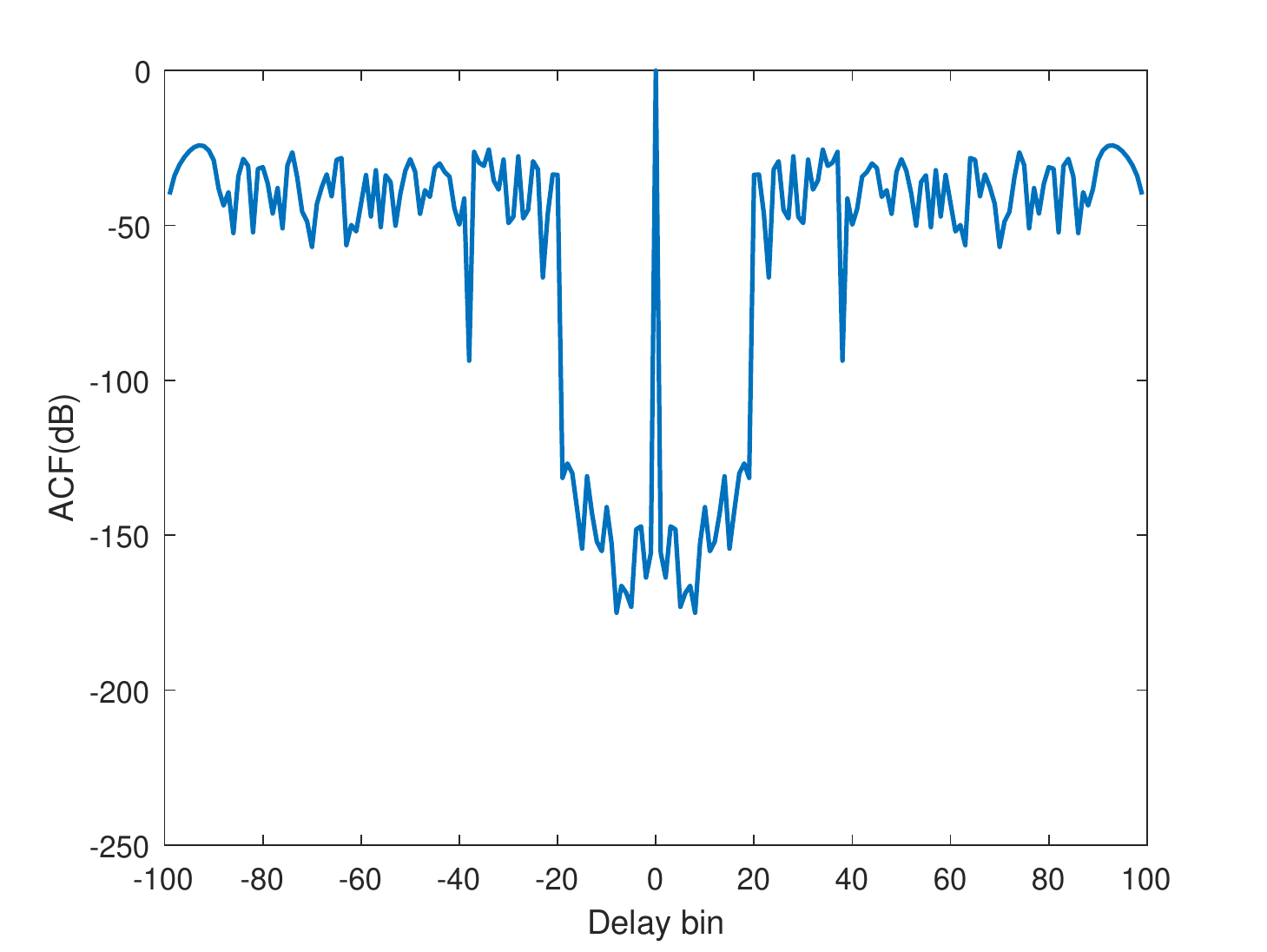}} \label{Fig:8a}} }
{\subfigure[]{{\includegraphics[width = 0.35\textwidth]{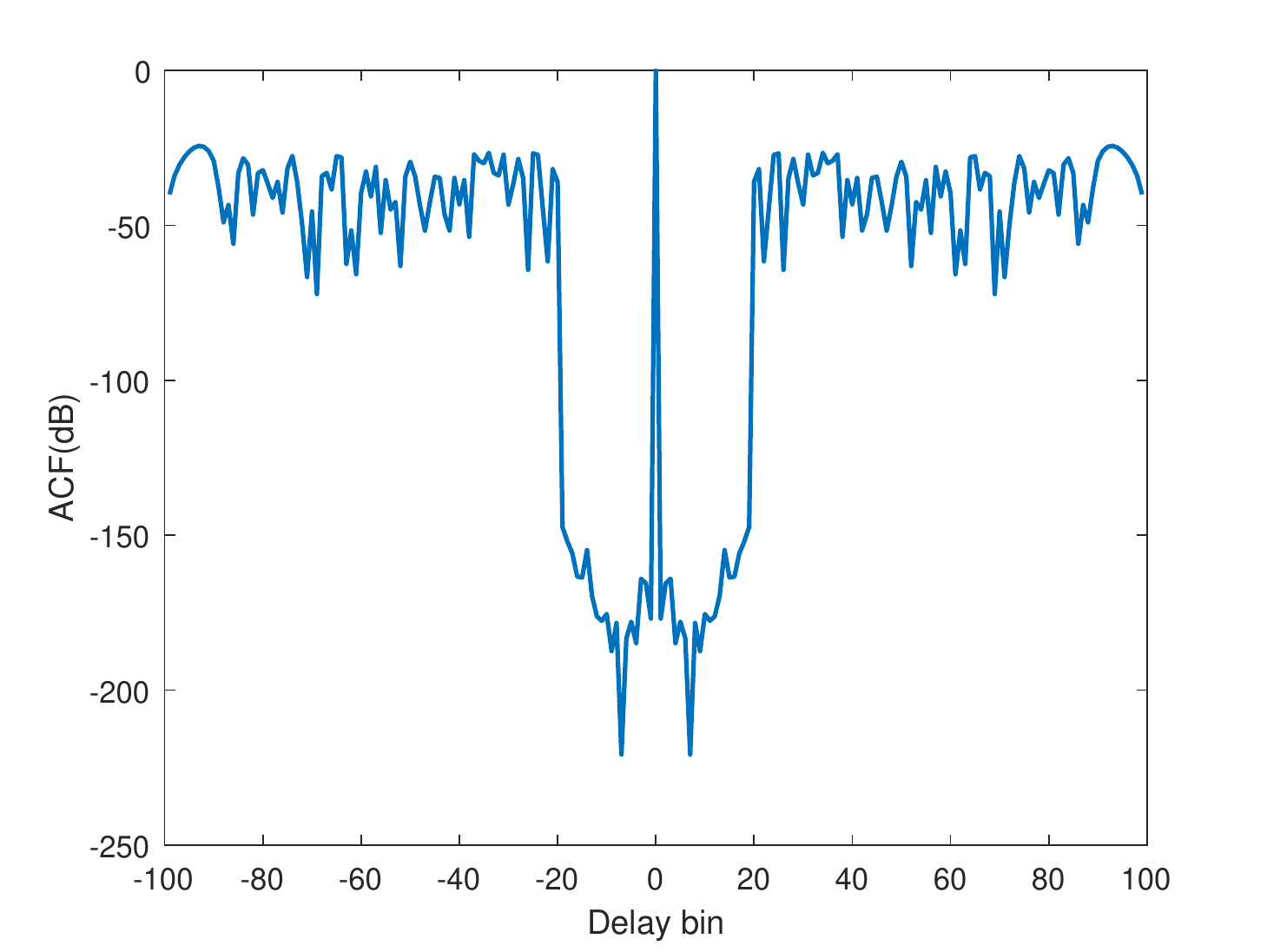}} \label{Fig:8b}} }
\caption{The ACF of the synthesized constant-modulus waveforms. $\bR_\txh = 0.1 \bI$. $\bR_\txn = \bI$. (a) Mutual information. (b) MMSE.}
\label{Fig:8}
\end{figure}

Finally, we show that if the interference in the received signal is dominated by the white noise, and the target covariance matrix is diagonal, the waveforms synthesized by the proposed algorithms will exhibit ZCZ properties. To this end, we set $\bR_\txh = 0.1 \bI$ and $\bR_\txn = \bI$. The convergence of mutual information and MMSE of the constant-modulus waveforms synthesized by the proposed algorithms are shown in Fig. \ref{Fig:7a} and Fig. \ref{Fig:7b}, respectively, where we initialize the proposed algorithms by using the LFM waveform, and we set $\varepsilon = 10^{-20}$. The curves in Fig. \ref{Fig:7} demonstrate that the objective values converge to the optimal values. Fig. \ref{Fig:8} plots the (normalized) auto-correlation function (ACF) of the waveforms synthesized by the proposed algorithms. We can observe that the waveform based on maximizing mutual information has very low correlation sidelobes (lower than about $-130$ dB) from $r_1$ to $r_{P-1}$. For waveform design based on minimizing MMSE, the proposed algorithm converges faster and achieves a lower sidelobe (lower than about $-150$ dB). Therefore, the synthesized waveforms in this situation have ZCZ properties. In other words, the ZCZ waveforms maximize the mutual information and minimize the MMSE.

\section{Conclusion} \label{sec:Conclusion}
This paper considered the design of practically constrained waveforms for radar range profiling. Two design metrics were used,  i.e., mutual information maximization and MMSE minimization. To tackle the non-convex waveform design problem, we developed an optimization framework based on MM.
Due to the ascent property of MM, the proposed algorithms had guaranteed convergence of objective values.
Finally, numerical examples demonstrated that the waveforms synthesized by the proposed algorithms were superior to the counterparts. %conventional LFM waveforms.

Possible future research track might concern the extension of the proposed algorithms to design waveforms for radar imaging. In addition, the design of practically constrained waveforms for target recognition/classification might be important (see, e.g., \cite{Gu2019ITGMM}, for a recent progress on this topic). Furthermore, the application of the proposed algorithms to polarimetric radar (see, e.g., \cite{Garren2002matchedillumination,Cheng2017PolarimetricDesign} for some discussions) is left for future work.

\appendices
\section{The Monotonicity of the design metric w.r.t. the transmit energy} \label{Apd:A}
In this appendix, we show that the mutual information is increasing with the waveform energy and the MMSE is decreasing with the waveform energy. To this end, we first prove $f_{{\textrm{I}}}(\alpha \bs)  \geq f_{{\textrm{I}}}(\bs)$, where $\alpha \geq 1$.
%\begin{align}%\label{}
%   f_{{\textrm{I}}}(\alpha \bs) & \geq f_{{\textrm{I}}}(\bs), \label{eq:MImono}%\\
%   %f_{{\textrm{E}}}(\alpha \bs) & \geq f_{{\textrm{E}}}(\bs), \label{eq:MSEmono}
%\end{align}
%where $\alpha \geq 1$.

We note that
\begin{equation*}%\label{}
  f_{{\textrm{I}}}(\alpha \bs) - f_{{\textrm{I}}}(\bs) = \log \det (\alpha^2 \bS  \bR_\txh \bS^\HH + \bR_\txn) - \log \det (\bS  \bR_\txh \bS^\HH + \bR_\txn).
\end{equation*}
Since $\alpha^2 \bS  \bR_\txh \bS^\HH \succeq \bS  \bR_\txh \bS^\HH$ and $\log\det(\bA)$ is an increasing function of $\bA \succ \bzero$, we have $f_{{\textrm{I}}}(\alpha \bs) - f_{{\textrm{I}}}(\bs) \geq 0$.
%\begin{equation}%\label{}
%  f_{{\textrm{I}}}(\alpha \bs) - f_{{\textrm{I}}}(\bs) \geq 0.
%\end{equation}

Next we prove that the MMSE is decreasing with the waveform energy. To this end, we rewrite the MMSE as
\begin{align}%\label{}
  &\tr (\bR_\txh - \bR_\txh\bS^\HH (\bS \bR_\txh \bS^\HH + \bR_\txn)^{-1} \bS\bR_\txh)\nonumber\\
  &=\tr\left((\bR_\txh^{-1} + \bS^\HH \bR_\txn^{-1}\bS)^{-1}\right).
\end{align}
Since $\alpha^2 \bS^\HH \bR_\txn^{-1}\bS \succeq \bS^\HH \bR_\txn^{-1}\bS$ and $\tr(\bA^{-1})$ is an decreasing function of $\bA \succ \bzero$, we can verify that the MMSE decreases with the transmit energy.

\section{Proof of Proposition \ref{Prop:1}} \label{Apd:Prop1}
First, it can be checked that  the $p$th column of $\bS$ can be written as $\bE_p \bs$. Thus,
\begin{align}\label{eq:vecS}
  \vec(\bS)
  &= [(\bE_1 \bs)^T,\cdots,(\bE_P \bs)^T]^T\nonumber\\
  &= [\bE_1^T,\cdots,\bE_P^T]^T \bs\nonumber\\
  &= \bE \bs.
\end{align}
  Using the fact that $\tr(\bA^H \bB) = \vec^\HH(\bA)\vec(\bB)$ \cite{bernstein2009matrix}, we have
  \begin{align}\label{eq:bk}
    \tr(\bS^\HH \bG_k^{21} \bR_\txh^{\frac{1}{2}}  )
    &= \vec^\HH(\bS) \vec(\bG_k^{21} \bR_\txh^{\frac{1}{2}} ) \nonumber\\
    &= \bs^\HH \bE^\HH \vec(\bG_k^{21} \bR_\txh^{\frac{1}{2}} )\nonumber\\
    &= \bs^\HH \ba_{\textrm{I},k}.
  \end{align}
In addition, noting that $\tr(\mathbf{ABCD}) = \vec^T(\bD)(\bA \otimes \bC^T) \vec(\bB^T)$ \cite{bernstein2009matrix}, we can obtain
\begin{align}\label{eq:Ak}
  \tr( \bG_k^{22} \bS \bR_\txh\bS^\HH)
  &= \tr( \bR_\txh\bS^\HH\bG_k^{22} \bS) \nonumber\\
  &= \vec^T(\bS) (\bR_\txh \otimes (\bG_k^{22})^T)\vec(\bS^*)\nonumber\\
  &= \bs^T \bE^T (\bR_\txh \otimes (\bG_k^{22})^T) \bE^* \bs^* \nonumber\\
  &= \bs^\HH \bE^\HH (\bR_\txh^* \otimes \bG_k^{22} ) \bE \bs \nonumber\\
  &= \bs^\HH \bA_{\textrm{I},k}\bs,
\end{align}
where the first line holds because $\tr(\bA\bB) = \tr(\bB\bA)$, the fourth line follows from that $\bG_k^{22}$ and $\bS \bR_\txh\bS^\HH$ are both Hermitian, and $\tr( \bG_k^{22} \bS \bR_\txh\bS^\HH) $ is real-valued.

 Combing \eqref{eq:bk} and \eqref{eq:Ak}, we complete the proof.

 \section{Proof of Lemma \ref{Lemma:2}} \label{Apd:Lemma1}
 According to \cite{Boyd2004convexBook}, the epigraph of $\psi(\bA,\bB)$ is defined by
    \begin{equation}\label{eq:epigramDef}
    \textrm{epi}(\psi) = \left\{ (\bA, \bB, t) |  \tr(\bA^\HH \bB^{-1} \bA) \leq t \right\}.
  \end{equation}
Note that $\tr(\bA^\HH \bB^{-1} \bA) \leq t$ is equivalent to $\{\tr(\bY) \leq t, \bA^\HH \bB^{-1} \bA \succeq \bY, \bY \succeq \bzero\}$. Thus, $\textrm{epi}(\psi)$ can be rewritten as
  \begin{equation*}%\label{eq:epigram}
    \textrm{epi}(\psi) = \left\{ (\bA, \bB, t) | \exists \bY \succeq \bzero, \begin{bmatrix}
                                                          \bB & \bA \\
                                                          \bA^\HH & \bY
                                                    \end{bmatrix}\succeq \bzero, \tr(\bY) \leq t \right\}.
  \end{equation*}
  It is easy to check that $\textrm{epi}(\psi)$ is convex. Therefore, $\psi(\bA,\bB)$ is a convex function of the matrix pair $(\bA,\bB)$ (namely, jointly convex). Using the fact that convex functions are minorized by their supporting hyperplanes, we have
  \begin{align*}%\label{eq:traceInvMM}
    \tr(\bA^\HH \bB^{-1} \bA)
    &\geq \tr(\bA_k^\HH \bB_k^{-1} \bA_k) + 2\Re(\tr(\bA_k^\HH \bB_k^{-1}(\bA-\bA_k)))\nonumber\\
     &\quad - \tr(\bB_k^{-1} \bA_k \bA_k^\HH \bB_k^{-1} (\bB-\bB_k))\nonumber\\
    &=2 \Re(\tr(\bA_k^\HH \bB_k^{-1}\bA)) - \tr(\bB_k^{-1} \bA_k \bA_k^\HH \bB_k^{-1} \bB),
  \end{align*}
  where $(\bA_k^\HH \bB_k^{-1}, -\bB_k^{-1} \bA_k \bA_k^\HH \bB_k^{-1})$ is the gradient of $\psi(\bA,\bB)$ at $(\bA_k,\bB_k)$ (which can be shown using the results from \cite{hjorungnes2007complex,Burke2015GMF}).

An alternative proof without using convexity is by noting that (since $\bB \succ \bzero$)
\begin{equation}%\label{}
  (\bB^{-1}\bA - \bB_k^{-1}\bA_k)^\HH \bB (\bB^{-1}\bA - \bB_k^{-1}\bA_k) \succeq \bzero.
\end{equation}
Thus,
\begin{equation}%\label{}
  \tr[(\bB^{-1}\bA - \bB_k^{-1}\bA_k)^\HH \bB (\bB^{-1}\bA - \bB_k^{-1}\bA_k)] \geq 0.
\end{equation}
After some algebraic manipulations, we complete the proof.%obtain \eqref{eq:traceInvMM}.
\section{Proof of negative semi-definiteness of $\bA_{\textrm{I},k}$ and $\bA_{\textrm{E},k}$} \label{Apd:B}
To prove that $\bA_{\textrm{I},k}$ is negative semi-definite, it is sufficient to prove that $\bR_\txh^* \otimes \bG_k^{22}$ is negative semi-definite. Note that $\bR_\txh \succ \bzero$. In addition, since $\bR_{\bs,k}\succeq \bR_\txn$, we have $\bG_k^{22} = \bR_{\bs,k}^{-1} - \bR_\txn^{-1} \preceq \bzero$. Thus, $\bR_\txh^* \otimes \bG_k^{22} \preceq \bzero$.
Similarly, to identify the negative semi-definiteness of $\bA_{\textrm{E},k}$, it is sufficient to prove that $\bR_\txh^* \otimes \bT_k \succeq \bzero$. Note that $\bT_k = \bR_{\bs,k}^{-1} \bS_k \bR_\txh^2 \bS_k^\HH \bR_{\bs,k}^{-1} \succeq \bzero$. Thus, $\bA_{\textrm{E},k}$ is negative semi-definite.

%\section*{Acknowledgment}

\ifCLASSOPTIONcaptionsoff
  \newpage
\fi

\bibliographystyle{IEEEtran}
\bibliography{IEEEabrv,hrrp}

\begin{thebibliography}{10}
\providecommand{\url}[1]{#1}
\csname url@samestyle\endcsname
\providecommand{\newblock}{\relax}
\providecommand{\bibinfo}[2]{#2}
\providecommand{\BIBentrySTDinterwordspacing}{\spaceskip=0pt\relax}
\providecommand{\BIBentryALTinterwordstretchfactor}{4}
\providecommand{\BIBentryALTinterwordspacing}{\spaceskip=\fontdimen2\font plus
\BIBentryALTinterwordstretchfactor\fontdimen3\font minus
  \fontdimen4\font\relax}
\providecommand{\BIBforeignlanguage}[2]{{%
\expandafter\ifx\csname l@#1\endcsname\relax
\typeout{** WARNING: IEEEtran.bst: No hyphenation pattern has been}%
\typeout{** loaded for the language `#1'. Using the pattern for}%
\typeout{** the default language instead.}%
\else
\language=\csname l@#1\endcsname
\fi
#2}}
\providecommand{\BIBdecl}{\relax}
\BIBdecl

\bibitem{Wehner1994highResolutionRadar}
D.~R. Wehner, \emph{High resolution radar}.\hskip 1em plus 0.5em minus
  0.4em\relax Norwood: Artech House, 1995.

\bibitem{tait2005introduction}
P.~Tait, \emph{Introduction to radar target recognition}.\hskip 1em plus 0.5em
  minus 0.4em\relax London: The Institution of Electronic Technology, 2005.

\bibitem{levanon2005cross}
N.~Levanon, ``Cross-correlation of long binary signals with longer mismatched
  filters,'' \emph{IEE Proceedings-Radar, Sonar and Navigation}, vol. 152,
  no.~6, pp. 377--382, 2005.

\bibitem{Stoica2008IV}
P.~Stoica, J.~Li, and M.~Xue, ``Transmit codes and receive filters for radar,''
  \emph{{IEEE} Signal Process. Mag.}, vol.~25, no.~6, pp. 94--109, 2008.

\bibitem{Song2000Estimation}
S.~M.~H. Song, W.~M. Kim, P.~Dongwook, and K.~Young-Sik, ``Estimation theoretic
  approach for radar pulse compression processing and its optimal codes,''
  \emph{Electronics Letters}, vol.~36, no.~3, pp. 250--252, 2000.

\bibitem{blunt2006APC}
S.~D. Blunt and K.~Gerlach, ``Adaptive pulse compression via {MMSE}
  estimation,'' \emph{{IEEE} Trans. Aerosp. Electron. Syst.}, vol.~42, no.~2,
  pp. 572--584, 2006.

\bibitem{Kikuchi2017APC}
H.~{Kikuchi}, E.~{Yoshikawa}, T.~{Ushio}, F.~{Mizutani}, and M.~{Wada},
  ``Adaptive pulse compression technique for {X-Band} phased array weather
  radar,'' \emph{IEEE Geoscience and Remote Sensing Letters}, vol.~14, no.~10,
  pp. 1810--1814, 2017.

\bibitem{yardibi2010IAA}
T.~Yardibi, J.~Li, P.~Stoica, M.~Xue, and A.~B. Baggeroer, ``Source
  localization and sensing: A nonparametric iterative adaptive approach based
  on weighted least squares,'' \emph{{IEEE} Trans. Aerosp. Electron. Syst.},
  vol.~46, no.~1, pp. 425--443, 2010.

\bibitem{He2012WaveformBook}
H.~He, J.~Li, and P.~Stoica, \emph{Waveform design for active sensing systems:
  a computational approach}.\hskip 1em plus 0.5em minus 0.4em\relax Cambridge
  University Press, 2012.

\bibitem{gini2012waveform}
F.~Gini, A.~De~Maio, and L.~Patton, \emph{Waveform design and diversity for
  advanced radar systems}.\hskip 1em plus 0.5em minus 0.4em\relax London:
  Institution of engineering and technology, 2012.

\bibitem{Cui2020Waveformbook}
G.~Cui, A.~De~Maio, A.~Farina, and J.~Li, \emph{Radar Waveform Design Based on
  Optimization Theory}.\hskip 1em plus 0.5em minus 0.4em\relax London: SciTech
  Publishing, 2020.

\bibitem{Stoica2009CAN}
P.~Stoica, H.~He, and J.~Li, ``New algorithms for designing unimodular
  sequences with good correlation properties,'' \emph{{IEEE} Trans. Signal
  Process.}, vol.~57, no.~4, pp. 1415--1425, 2009.

\bibitem{Song2015MM}
J.~Song, P.~Babu, and D.~P. Palomar, ``Optimization methods for designing
  sequences with low autocorrelation sidelobes,'' \emph{{IEEE} Trans. Signal
  Process.}, vol.~63, no.~15, pp. 3998--4009, 2015.

\bibitem{Garren2002matchedillumination}
D.~A. Garren, A.~Odom, M.~Osborn, J.~S. Goldstein, S.~Pillai, and J.~Guerci,
  ``Full-polarization matched-illumination for target detection and
  identification,'' \emph{{IEEE} Trans. Aerosp. Electron. Syst.}, vol.~38,
  no.~3, pp. 824--837, 2002.

\bibitem{Cheng2017PolarimetricDesign}
X.~Cheng, A.~Aubry, D.~Ciuonzo, A.~D. Maio, and X.~Wang, ``Robust waveform and
  filter bank design of polarimetric radar,'' \emph{{IEEE} Trans. Aerosp.
  Electron. Syst.}, vol.~53, no.~1, pp. 370--384, 2017.

\bibitem{Ciuonzo2015Intrapulse}
D.~Ciuonzo, A.~De~Maio, G.~Foglia, and M.~Piezzo, ``Intrapulse radar-embedded
  communications via multiobjective optimization,'' \emph{{IEEE} Trans. Aerosp.
  Electron. Syst.}, vol.~51, no.~4, pp. 2960--2974, 2015.

\bibitem{my2016TxRx}
B.~Tang and J.~Tang, ``Joint design of transmit waveforms and receive filters
  for {MIMO} radar space-time adaptive processing,'' \emph{{IEEE} Trans. Signal
  Process.}, vol.~64, no.~18, pp. 4707--4722, 2016.

\bibitem{my2020Polyphase}
B.~Tang, J.~Tuck, and P.~Stoica, ``Polyphase waveform design for {MIMO} radar
  space time adaptive processing,'' \emph{{IEEE} Trans. Signal Process.},
  vol.~68, pp. 2143--2154, 2020.

\bibitem{Bell1993IT}
M.~R. Bell, ``Information theory and radar waveform design,'' \emph{{IEEE}
  Trans. Inf. Theory}, vol.~39, no.~5, pp. 1578--1597, 1993.

\bibitem{Leshem2007Information}
A.~Leshem, O.~Naparstek, and A.~Nehorai, ``Information theoretic adaptive radar
  waveform design for multiple extended targets,'' \emph{IEEE Journal of
  Selected Topics in Signal Processing}, vol.~1, no.~1, pp. 42--55, 2007.

\bibitem{Romero2011MI}
R.~A. Romero, J.~Bae, and N.~A. Goodman, ``Theory and application of {SNR} and
  mutual information matched illumination waveforms,'' \emph{{IEEE} Trans.
  Aerosp. Electron. Syst.}, vol.~47, no.~2, pp. 912--927, 2011.

\bibitem{Huang2015spectrum}
K.-W. Huang, M.~Bic\v{a}, U.~Mitra, and V.~Koivunen, ``Radar waveform design in
  spectrum sharing environment: Coexistence and cognition,'' in \emph{IEEE
  Radar Conference (RadarCon)}, 2015, pp. 1698--1703.

\bibitem{Bica2016MI}
M.~Bic\v{a}, K.-W. Huang, V.~Koivunen, and U.~Mitra, ``Mutual information based
  radar waveform design for joint radar and cellular communication systems,''
  in \emph{IEEE International Conference on Acoustics, Speech and Signal
  Processing (ICASSP)}.\hskip 1em plus 0.5em minus 0.4em\relax IEEE, 2016, pp.
  3671--3675.

\bibitem{Palama2019matched}
R.~Palam\'{a}, H.~Griffiths, and F.~Watson, ``Joint dynamic spectrum access and
  target-matched illumination for cognitive radar,'' \emph{IET Radar, Sonar \&
  Navigation}, vol.~13, no.~5, pp. 750--759, 2019.

\bibitem{my2019Spectrally}
B.~Tang and J.~Li, ``Spectrally constrained {MIMO} radar waveform design based
  on mutual information,'' \emph{{IEEE} Trans. Signal Process.}, vol.~67,
  no.~3, pp. 821--834, 2019.

\bibitem{Karbasi2015ExtendedTargets}
S.~M. Karbasi, A.~Aubry, A.~De~Maio, and M.~H. Bastani, ``Robust transmit code
  and receive filter design for extended targets in clutter,'' \emph{{IEEE}
  Trans. Signal Process.}, vol.~63, no.~8, pp. 1965--1976, 2015.

\bibitem{my2016ICASSP}
B.~Tang and J.~Tang, ``Robust waveform design of wideband cognitive radar for
  extended target detection,'' in \emph{IEEE International Conference on
  Acoustics, Speech and Signal Processing (ICASSP)}, 2016, Conference
  Proceedings, pp. 3096--3100.

\bibitem{Gurbuz2019CognitiveOverview}
S.~Z. Gurbuz, H.~D. Griffiths, A.~Charlish, M.~Rangaswamy, M.~S. Greco, and
  K.~Bell, ``An overview of cognitive radar: Past, present, and future,''
  \emph{{IEEE} Aerosp. Electron. Syst. Mag.}, vol.~34, no.~12, pp. 6--18, 2019.

\bibitem{Horne2020AdaptiveSignalling}
C.~P. Horne, A.~M. Jones, G.~E. Smith, and H.~D. Griffiths, ``Fast fully
  adaptive signalling for target matching,'' \emph{{IEEE} Aerosp. Electron.
  Syst. Mag.}, vol.~35, no.~6, pp. 46--62, 2020.

\bibitem{Smith2016cognitiveExperiments}
G.~E. Smith, Z.~Cammenga, A.~Mitchell, K.~L. Bell, J.~Johnson, M.~Rangaswamy,
  and C.~Baker, ``Experiments with cognitive radar,'' \emph{{IEEE} Aerosp.
  Electron. Syst. Mag.}, vol.~31, no.~12, pp. 34--46, 2016.

\bibitem{cover2012elements}
T.~M. Cover and J.~A. Thomas, \emph{Elements of information theory}.\hskip 1em
  plus 0.5em minus 0.4em\relax New Jersey: John Wiley \& Sons, 2012.

\bibitem{kaybook1993}
S.~M. Kay, \emph{Fundamentals of Statistical Signal Processing: Estimation
  Theory}.\hskip 1em plus 0.5em minus 0.4em\relax New Jersey: Prentice Hall,
  1993.

\bibitem{HornJohnson1990matrixbook}
R.~A. Horn and C.~R. Johnson, \emph{Matrix analysis}.\hskip 1em plus 0.5em
  minus 0.4em\relax Cambridge: Cambridge University Press, 1990.

\bibitem{DeMaio2009Similarity}
A.~De~Maio, S.~De~Nicola, Y.~Huang, Z.-Q. Luo, and S.~Zhang, ``Design of phase
  codes for radar performance optimization with a similarity constraint,''
  \emph{{IEEE} Trans. Signal Process.}, vol.~57, no.~2, pp. 610--621, 2009.

\bibitem{DeMaio2011PAPR}
A.~De~Maio, Y.~Huang, M.~Piezzo, S.~Zhang, and A.~Farina, ``Design of optimized
  radar codes with a peak to average power ratio constraint,'' \emph{{IEEE}
  Trans. Signal Process.}, vol.~59, no.~6, pp. 2683--2697, 2011.

\bibitem{Stoica2012joint}
P.~Stoica, H.~He, and J.~Li, ``Optimization of the receive filter and transmit
  sequence for active sensing,'' \emph{{IEEE} Trans. Signal Process.}, vol.~60,
  no.~4, pp. 1730--1740, 2012.

\bibitem{Soltanalian2013crew}
M.~Soltanalian, B.~Tang, J.~Li, and P.~Stoica, ``Joint design of the receive
  filter and transmit sequence for active sensing,'' \emph{IEEE Signal
  Processing Letters}, vol.~20, no.~5, pp. 423--426, 2013.

\bibitem{Wu2018MultipleConstraints}
L.~Wu, P.~Babu, and D.~P. Palomar, ``Transmit waveform/receive filter design
  for {MIMO} radar with multiple waveform constraints,'' \emph{{IEEE} Trans.
  Signal Process.}, vol.~66, no.~6, pp. 1526--1540, 2018.

\bibitem{Aubry2020MultiSpectral}
A.~Aubry, A.~D. Maio, M.~A. Govoni, and L.~Martino, ``On the design of
  multi-spectrally constrained constant modulus radar signals,'' \emph{{IEEE}
  Trans. Signal Process.}, vol.~68, pp. 2231--2243, 2020.

\bibitem{my2021RangeSpread}
B.~Tang and P.~Stoica, ``Information-theoretic waveform design for {MIMO} radar
  detection in range-spread clutter,'' \emph{Signal Processing}, vol. 182, p.
  107961, 2021.

\bibitem{Griffiths2015spectrum}
H.~Griffiths, L.~Cohen, S.~Watts, E.~Mokole, C.~Baker, M.~Wicks, and S.~Blunt,
  ``Radar spectrum engineering and management: technical and regulatory
  issues,'' \emph{Proceedings of the IEEE}, vol. 103, no.~1, pp. 85--102, 2015.

\bibitem{Bockmair2019CognitivePrinciples}
M.~Bockmair, C.~Fischer, M.~Letsche-Nuesseler, C.~Neumann, M.~Schikorr, and
  M.~Steck, ``Cognitive radar principles for defence and security
  applications,'' \emph{{IEEE} Aerosp. Electron. Syst. Mag.}, vol.~34, no.~12,
  pp. 20--29, 2019.

\bibitem{Carotenuto2020SpectrumManagement}
V.~Carotenuto, A.~Aubry, A.~D. Maio, N.~Pasquino, and A.~Farina, ``Assessing
  agile spectrum management for cognitive radar on measured data,''
  \emph{{IEEE} Aerosp. Electron. Syst. Mag.}, vol.~35, no.~6, pp. 20--32, 2020.

\bibitem{yang2020MultiSpectral}
J.~Yang, A.~Aubry, A.~De~Maio, X.~Yu, and G.~Cui, ``Design of constant modulus
  discrete phase radar waveforms subject to multi-spectral constraints,''
  \emph{IEEE Signal Processing Letters}, vol.~27, pp. 875--879, 2020.

\bibitem{Aubry2014spectrally}
A.~Aubry, A.~De~Maio, M.~Piezzo, and A.~Farina, ``Radar waveform design in a
  spectrally crowded environment via nonconvex quadratic optimization,''
  \emph{{IEEE} Trans. Aerosp. Electron. Syst.}, vol.~50, no.~2, pp. 1138--1152,
  2014.

\bibitem{Aubry2016spectrally}
A.~Aubry, V.~Carotenuto, A.~De~Maio, A.~Farina, and L.~Pallotta, ``Optimization
  theory-based radar waveform design for spectrally dense environments,''
  \emph{{IEEE} Aerosp. Electron. Syst. Mag.}, vol.~31, no.~12, pp. 14--25,
  2016.

\bibitem{Hunter2004MM}
D.~R. Hunter and K.~Lange, ``A tutorial on {MM} algorithms,'' \emph{The
  American Statistician}, vol.~58, no.~1, pp. 30--37, 2004.

\bibitem{Sun2017MM}
Y.~Sun, P.~Babu, and D.~P. Palomar, ``Majorization-minimization algorithms in
  signal processing, communications, and machine learning,'' \emph{{IEEE}
  Trans. Signal Process.}, vol.~65, no.~3, pp. 794--816, 2017.

\bibitem{my2018Efficient}
B.~Tang, Y.~Zhang, and J.~Tang, ``An efficient minorization maximization
  approach for {MIMO} radar waveform optimization via relative entropy,''
  \emph{{IEEE} Trans. Signal Process.}, vol.~66, no.~2, pp. 400--411, 2018.

\bibitem{Boyd2004convexBook}
S.~Boyd and L.~Vandenberghe, \emph{Convex optimization}.\hskip 1em plus 0.5em
  minus 0.4em\relax Cambridge: Cambridge University Press, 2004.

\bibitem{hjorungnes2007complex}
A.~Hjorungnes and D.~Gesbert, ``Complex-valued matrix differentiation:
  Techniques and key results,'' \emph{{IEEE} Trans. Signal Process.}, vol.~55,
  no.~6, pp. 2740--2746, 2007.

\bibitem{Tropp2005AP}
J.~A. Tropp, I.~S. Dhillon, R.~W. Heath, and T.~Strohmer, ``Designing
  structured tight frames via an alternating projection method,'' \emph{{IEEE}
  Trans. Inf. Theory}, vol.~51, no.~1, pp. 188--209, 2005.

\bibitem{Ai2011rankone}
W.~Ai, Y.~Huang, and S.~Zhang, ``New results on hermitian matrix rank-one
  decomposition,'' \emph{Mathematical programming}, vol. 128, no. 1-2, pp.
  253--283, 2011.

\bibitem{Boyd2011ADMM}
S.~Boyd, N.~Parikh, E.~Chu, B.~Peleato, and J.~Eckstein, ``Distributed
  optimization and statistical learning via the alternating direction method of
  multipliers,'' \emph{Foundations and Trends® in Machine Learning}, vol.~3,
  no.~1, pp. 1--122, 2011.

\bibitem{my2018admm}
B.~Tang, J.~Li, and J.~Liang, ``Alternating direction method of multipliers for
  radar waveform design in spectrally crowded environments,'' \emph{Signal
  Processing}, vol. 142, pp. 398--402, 2018.

\bibitem{Li2006SWORD}
J.~Li, J.~R. Guerci, and L.~Xu, ``Signal waveform's optimal-under-restriction
  design for active sensing,'' \emph{IEEE Signal Processing Letters}, vol.~13,
  no.~9, pp. 565--568, 2006.

\bibitem{fan1999ZCZ}
P.~Z. Fan, N.~Suehiro, N.~Kuroyanagi, and X.~M. Deng, ``Class of binary
  sequences with zero correlation zone,'' \emph{Electronics Letters}, vol.~35,
  no.~10, pp. 777--779, 1999.

\bibitem{Yang2007MMSE}
Y.~Yang and R.~S. Blum, ``{MIMO} radar waveform design based on mutual
  information and minimum mean-square error estimation,'' \emph{{IEEE} Trans.
  Aerosp. Electron. Syst.}, vol.~43, no.~1, pp. 330--343, 2007.

\bibitem{Gu2019ITGMM}
Y.~Gu and N.~A. Goodman, ``Information-theoretic waveform design for {Gaussian}
  mixture radar target profiling,'' \emph{{IEEE} Trans. Aerosp. Electron.
  Syst.}, vol.~55, no.~3, pp. 1528--1536, 2019.

\bibitem{bernstein2009matrix}
D.~S. Bernstein, \emph{Matrix mathematics: theory, facts, and formulas}.\hskip
  1em plus 0.5em minus 0.4em\relax Princeton University Press, 2009.

\bibitem{Burke2015GMF}
J.~V. Burke and T.~Hoheisel, ``Matrix support functionals for inverse problems,
  regularization, and learning,'' \emph{SIAM Journal on Optimization}, vol.~25,
  no.~2, pp. 1135--1159, 2015.

\end{thebibliography}
\end{document}